\documentclass[aps,showpacs,nofootinbib,showkeys,11pt]{revtex4-1}
\usepackage{blindtext}
\usepackage[utf8]{inputenc}
\usepackage{chngcntr}
\counterwithout{section}{equation}
\usepackage{amsmath}
\usepackage{amsfonts}
\usepackage{amssymb}
\usepackage{ragged2e}
\usepackage{makeidx}
\usepackage{enumerate}
\usepackage{hhline}
\usepackage{array}
\usepackage{tabularx}
\usepackage{graphicx}
\def\beq{\begin{equation}}
\def\eeq{\end{equation}}
\def\bea{\begin{eqnarray}}
\def\eea{\end{eqnarray}}

\usepackage{mathtools}
\allowdisplaybreaks[4]
\usepackage[unicode]{hyperref}
\hypersetup{
     colorlinks   = true,
     citecolor    = red,
     linkcolor    = blue,
     urlcolor     = cyan,
}

\begin{document}

\tolerance=5000

\title{Effective Theory of Inflationary Magnetogenesis and Constraints on Reheating}
\author{Debaprasad~Maity}
\email{debu@iitg.ac.in}
\affiliation{Department of Physics, Indian Institute of Technology Guwahati, Assam 781039, India}                
\author{Sourav~Pal}
\email{pal.sourav@iitg.ac.in}
\affiliation{Department of Physics, Indian Institute of Technology Guwahati, Assam 781039, India}
\author{Tanmoy~Paul$^{(1,2)}$}
\email{pul.tnmy9@gmail.com}
\affiliation{$^{(1)}$~Department of Physics, Chandernagore College, Hooghly - 712 136, India.\\
$^{(2)}$~Laboratory for Theoretical Cosmology, TUSUR, 634050 Tomsk, Russia.}
\date{\today}
\tolerance=5000

\begin{abstract}
Effective theory framework based on symmetry has recently gained widespread interest in the field of cosmology. 
In this paper, we apply the same idea on the genesis of the primordial magnetic field and its evolution throughout the 
cosmological universe. Given the broken time-diffeomorphism symmetry by the cosmological background, we considered the 
most general Lagrangian of electromagnetic and metric fluctuation up to second order, which naturally breaks conformal 
symmetry in the electromagnetic (EM) sector. We also include parity violation in the electromagnetic sector 
with the motivation that has potential observational significance. In such a set-up, we explore the evolution of EM, scalar, 
and tensor perturbations considering different observational constraints. In our analysis we emphasize the role played by the 
intermediate reheating phase which has got limited interest in all the previous studies. Assuming the vanishing electrical 
conductivity during the entire period of reheating, the well-known Faraday electromagnetic induction has been shown to play a 
crucial role in enhancing the strength of the present-day magnetic field. We show how such physical effects combined with the PLANCK 
and the large scale magnetic field observation makes a large class of models viable and severely restricts the reheating equation 
of state parameter within a very narrow range of $0.01 < \omega_\mathrm{eff} < 0.27$, 
which is nearly independent of reheating scenarios we have considered.

\end{abstract}


\maketitle

\section{Introduction}

Magnetic fields are observed over a wide range of scales in our universe. They have been detected in intergalactic voids, within galaxy clusters and even 
in individual galaxies \cite{Grasso:2000wj,Beck:2000dc,Widrow:2002ud,Kandus:2010nw,Durrer:2013pga,Subramanian:2015lua}. Physical mechanism of the origin of 
such magnetic fields across wide range of scales has not been completely understood. 
There are two approaches which have been widely discussed in the literature. Specifically for the magnetic field at the astrophysical scale, 
physical processes such as Biermann battery \cite{Biermann} 
play crucial role in providing seed fields which are subsequently amplified by dynamo mechanism in the 
plasma \cite{Kulsrud:2007an,Brandenburg:2004jv,Subramanian:2009fu} .The second approach deals with the primordial generation of 
seed magnetic fields during inflationary phase \cite{Sharma:2017eps,Sharma:2018kgs,Jain:2012ga,Durrer:2010mq,Kanno:2009ei,Campanelli:2008kh,
Demozzi:2009fu,Bamba:2008ja,Bamba:2008xa,Bamba:2012mi,Bamba:2006ga,Bamba:2003av,Bamba:2004cu,Bamba:2020qdj,Haque:2020bip,Giovannini:2020zjo,
Giovannini:2017rbc,Giovannini:2003yn,Kobayashi:2019uqs,Ratra:1991bn,Ade:2015cva,Chowdhury:2018mhj,Vachaspati:1991nm,Turner:1987bw,
Takahashi:2005nd,Agullo:2013tba,Ferreira:2013sqa,Atmjeet:2014cxa,Kushwaha:2020nfa,Sharma:2021rot,Adshead:2015pva,Adshead:2016iae} , 
during contracting phase \cite{Frion:2020bxc,Chowdhury:2016aet,Chowdhury:2018blx,Koley:2016jdw,Qian:2016lbf,Membiela:2013cea}. 
A confirmed detection of magnetic fields in the large voids in our universe might in principle indicate their primordial origin.

Among all the proposals so far, inflationary magnetogenesis has earned a lot of attention due its simplicity and elegance.  
Apart from solving flatness and horizon problems of standard big-bang, inflation predicts a nearly scale invariant power spectrum 
of CMB and matter density distribution that are consistent with the Planck observations with great precision
\cite{guth,Linde:2005ht,Langlois:2004de,Riotto:2002yw,Baumann:2009ds,Bamba:2015uma}. Therefore, same paradigm playing the role in generating large scale magnetic field would have larger theoretical motivation. 
The standard Maxwell's theory  is conformal invariant. Therefore, magnetic field can not be generated in the conformally flat inflationary background. 
The simplest way to harness electromagnetic energy from the background inflaton energy, therefore, is to break this conformal invariance. Several 
magnetogenesis models have been proposed along this line where the conformal 
invariance is broken by introducing explicit coupling between EM field with inflaton, axion, or higher curvature term.  
\cite{Sharma:2017eps,Sharma:2018kgs,Jain:2012ga,Durrer:2010mq,Kanno:2009ei,Campanelli:2008kh,
Demozzi:2009fu,Bamba:2008ja,Bamba:2008xa,Bamba:2012mi,Bamba:2006ga,Bamba:2003av,Bamba:2004cu,Bamba:2020qdj,Haque:2020bip,Giovannini:2020zjo,
Giovannini:2017rbc,Giovannini:2003yn,Kobayashi:2019uqs,Ratra:1991bn,Ade:2015cva,
Chowdhury:2018mhj,Vachaspati:1991nm,Turner:1987bw,Takahashi:2005nd,Agullo:2013tba,Ferreira:2013sqa,Atmjeet:2014cxa,Kushwaha:2020nfa,Sharma:2021rot,
Adshead:2015pva,Adshead:2016iae,Caprini:2014mja,Kobayashi:2014sga,Atmjeet:2013yta,Fujita:2015iga,Campanelli:2015jfa,Tasinato:2014fia}. 
However this simple mechanism of  inflationary 
magnetogenesis is riddled with some obvious theoretical problems. Those are related to backreaction and the strong coupling problems. The 
backreaction problem arises when the 
EM field energy density overshoots the background energy density, by which the inflationary evolution of the scale factor may be spoiled. 
On other hand, the strong coupling problem occurs when the effective electric charge becomes strong during inflation, which makes the 
perturbative calculation questionable. Thus in order to ensure the viability of an inflationary magnetogenesis model, the 
backreaction and the strong coupling issues should be  addressed and resolved in the model 
(see \cite{Sharma:2017eps,Demozzi:2009fu,Ferreira:2013sqa,Tasinato:2014fia}).

Apart from the inflationary perspective, the magnetic field generation in the context of bouncing cosmology has also 
been proposed earlier in the literature \cite{Frion:2020bxc,Chowdhury:2016aet,Chowdhury:2018blx,Koley:2016jdw,Qian:2016lbf,Membiela:2013cea}. 
However, it may be mentioned that in spite of predicting scale 
invariant power spectrum(s) consistent with the Planck observations, the bouncing model(s), generally,  
suffers from important theoretical issues such as violation of energy condition at the bounce point, the BKL 
instability associated with the growth of anisotropy at the contracting stage of the universe, the instability of the scalar and tensor perturbations etc 
\cite{Brandenberger:2012zb,Brandenberger:2016vhg,Battefeld:2014uga,Novello:2008ra,Cai:2014bea,Nojiri:2019lqw,Odintsov:2015ynk}. Such problems can be 
rescued to some extend in some modified theories of gravity including the higher curvature theories or extra dimensional model 
\cite{Battefeld:2014uga,Cai:2008qw,Elizalde:2019tee,Elizalde:2020zcb,Navo:2020eqt,Bamba:2014mya,Odintsov:2020zct,Banerjee:2020uil}. 

The Effective Field Theory (EFT) of cosmology is extremely powerful 
and has been widely used to study inflation \cite{Cheung:2007st,Weinberg:2008hq,Qiu:2020qsq}, bounce \cite{Cai:2016thi,Cai:2017tku} 
and dark energy \cite{Gubitosi:2012hu,Gleyzes:2013ooa,Piazza:2013coa}. 
Motivated by these works, in the present paper, we explore the inflationary 
magnetogenesis from the EFT perspective. Our basic framework would be effective theory of inflationary fluctuations coupled 
with the electromagnetic field. Time dependent inflaton background naturally breaks time diffeomorphism keeping spatial diffeomorphism intact. 
Using this spatial diffeomorphism symmetry, most general effective action for the metric perturbation coupled with the EM field can be written. 
Apart from the coupling with the spacetime perturbation, the EM field action also has self interaction terms which behave 
as scalar quantity under the spatial 
diffeomorphism symmetry transformation. 
Such self interaction couplings, along with that with the spacetime perturbation, spoil the conformal invariance in the 
EM sector and eventually leads to the gauge field production from primordial vacuum. 
In such a set-up, we have explored the evolution of the EM, scalar and tensor perturbation fields throughout cosmological 
evolution starting from the inflationary era. 
In this regard, we discuss two different scenarios depending on whether the magnetic power spectrum or the electric power 
spectrum become scale invariant in the inflationary superhorizon scale. After the inflationary epoch, 
the universe enters into a reheating era, and based on the reheating dynamics, 
we consider two different cases: (i) Firstly we assume an instantaneous reheating scenario 
where the universe makes a sudden jump from the inflationary 
epoch to the radiation dominated epoch, (ii) In the second case we consider the universe undergoing a reheating phase with non-zero e-fold number. 
In particular, we consider the conventional reheating mechanism where the inflaton field instantaneously converts to radiation energy density 
at the end of reheating, as proposed by Kamionkowski et al. \cite{Dai:2014jja} (see also \cite{Cook:2015vqa}). 
In such scenario, the main idea is to parameterize the reheating phase by a constant effective equation of state. 
(iii) Finally we consider  perturbative reheating scenario where the inflaton continuously decays into radiation and thus the effective equation of state during reheating becomes time dependent \cite{Albrecht:1982mp,Ellis:2015pla,Ueno:2016dim,Eshaghi:2016kne,Maity:2018qhi,Haque:2020zco,Haque:2019prw,Maity:2019ltu,Maity:2018exj,Maity:2018dgy,Maity:2016uyn,Bhattacharjee:2016ohe,DiMarco:2017zek,Drewes:2017fmn,DiMarco:2018bnw}. 
Because of the qualitative differences in the aforementioned two reheating scenarios, quantitatively different constraints on the effective theory as well as reheating parameters can be observed.  
For example, the presence of the reheating phase with a non-zero e-fold number has been shown to enhance the strength of the magnetic field as opposed to the instantaneous reheating case. Considering both CMB and large scale magnetic field observations this will naturally put constraints on the model parameters depending on the  reheating mechanisms.


The paper is organized as follows: after describing the model in Sec.\ref{sec_model}, we give the general expressions for the 
power spectra of EM, scalar and tensor perturbation fields in the present context in Sec.\ref{sec_power spectra}. 
The solution for the vector potential and the metric perturbation variables during inflation are presented in Sec.\ref{sec_solution_inflation}. 
The qualitative features of two different scenarios depending on whether the magnetic power spectrum or the electric power spectrum become scale invariant 
are carried out in Sec.\ref{sec_case-I} and Sec.\ref{sec_case-II} respectively, considering conventional instantaneous reheating model. 
In Sec.\ref{sec_reheating}, the present value of the magnetic field has been calculated considering conventional reheating phase wherein two possbilities with constant and time dependent reheating equation of state have been considered separately. Finally we summarise our results with some future works to be done.

\section{The model}\label{sec_model}
We start with the following effective field theory action,
\begin{eqnarray}
S = S_{bg} + S_{em} + S_{sp} + S_{int} ,
\label{full action0}
\end{eqnarray}
with $S_{bg}$ denotes the background action, $S_{em}$ is the electromagnetic field action, $S_{sp}$ symbolizes the action of the metric 
perturbation and $S_{int}$ represents the interaction between the the electromagnetic field and the metric perturbation variable. Following idea of 
effective field theory (EFT) inflation \cite{Cheung:2007st,Weinberg:2008hq}, the background action is expressed as,
\begin{eqnarray}
 S_{bg} = \int d^4x\sqrt{-g}\bigg[\frac{1}{2}M_\mathrm{Pl}^2R - \Lambda(\eta) - c(\eta)g^{00}\bigg]~~~.
 \label{background action}
\end{eqnarray}
where $M_\mathrm{Pl}$ is the Planck mass and $R$ is the Ricci scalar. Moreover, 
$f(\eta)$, $\Lambda(\eta)$ and $c(\eta)$ are the background EFT parameters and can be fixed by the background equations. The metric component $g^{00}$ is one of the scalars under spacial diffeomophism symmetry $x^i = x^i + \xi^i(x,t)$, where $\xi^i$ is arbitray spatial vector. For our present purpose we consider a spatially flat FLRW background metric,
\begin{eqnarray}
 ds^2 = -dt^2 + a^2(t)  \delta_{ij}dx^idx^j = a^2(\eta)(-d\eta^2 + \delta_{ij}dx^idx^j)~~.
 \label{FRW metric ansatz}
\end{eqnarray}
with $\eta$ being the conformal time and related to the cosmic time ($t$) as $\eta = \int {dt}/{a(t)}$. 
The above FRW metric along with the action (\ref{background action}) immediately lead to the background Friedmann equations as,
\begin{eqnarray}
 3M_\mathrm{Pl}^2\mathcal{H}^2 = a^2(\eta)\bigg(c(\eta) + \Lambda(\eta)\bigg)\nonumber\\
 -M_\mathrm{Pl}^2\bigg(2\mathcal{H}' + \mathcal{H}^2\bigg) = 
 a^2(\eta)\bigg(c(\eta) - \Lambda(\eta)\bigg) ,
 \label{background Freidmann equations}
\end{eqnarray}
where $\mathcal{H} = {a'}/{a}$ is known as the conformal Hubble parameter which is related to the cosmic Hubble parameter 
($H = \dot{a}/a$) as: $\mathcal{H} = aH$ (from now onwards, a prime represents $\frac{d}{d\eta}$ and an overdot symbolizes $\frac{d}{dt}$). 
Here it may be interesting to present few modified gravity models and its mapping with model independent EFT framework (Eq.\ref{background action}). Comparing the Friedmann Eqs.(\ref{background Freidmann equations}) with that of a specific gravity model, one can get certain forms of the functions $c(\eta)$ and $\Lambda(\eta)$ correspond to the respective model. 
For example, scalar-Einstein-Gauss-Bonnet gravity theory, which 
can be consistent with Planck results for suitable choices of the Gauss-Bonnet 
coupling function and the scalar field potential \cite{Li:2007jm,Odintsov:2018nch,Carter:2005fu,Nojiri:2019dwl,Elizalde:2010jx,Makarenko:2016jsy,delaCruzDombriz:2011wn,Bamba:2007ef,Chakraborty:2018scm,
Kanti:2015pda,Kanti:2015dra,Odintsov:2018zhw,Saridakis:2017rdo,Cognola:2006eg}, is described by the following time dependent effective theory 
parameters, \begin{eqnarray}
c(\eta(t))&=&\frac{1}{2}\dot{\phi}^2 - 4H^2\ddot{h} - 8H\dot{H}\dot{h} + 4H^3\dot{h}\nonumber\\
\Lambda(\eta(t))&=&V(\phi) + 4H^2\ddot{h} + 8H\dot{H}\dot{h} + 20H^3\dot{h}
 \label{embed15}
\end{eqnarray}
where $\phi$ is the scalar field under consideration (generally  the inflaton field), 
$V(\phi)$ and $h(\phi)$ denote the scalar field potential and the Gauss-Bonnet coupling function (with the scalar field) respectively. 
Thereby the scalar-Einstein-Gauss-Bonnet model is mapped into the EFT action (\ref{background action}) 
with the above expression for $c(\eta)$ and $\Lambda(\eta)$. Similarly the $F(R)$ gravity 
theory \cite{Nojiri:2010wj,Nojiri:2017ncd,Capozziello:2011et,Elizalde:2018rmz} can be embedded within the EFT action 
(\ref{background action}) for the following forms of $c(\eta)$ and $\Lambda(\eta)$ as,
\begin{eqnarray}
 c(\eta(t))&=&\dot{H}f'(R) + 3\big(\dddot{H} + 4\dot{H}^2 + 3H\dot{H} - 4H^2\dot{H}\big)f''(R) + 18\big(4H\dot{H} + \ddot{H}\big)^2f'''(R)\\
 \Lambda(\eta(t))&=&-\frac{f(R)}{2} + \big(2\dot{H} + 3H^2\big)f'(R) - 3\big(\dddot{H} + 4\dot{H}^2 + 9H\dot{H} + 20H^2\dot{H}\big)f''(R) 
 - 18\big(4H\dot{H} + \ddot{H}\big)^2f'''(R) \nonumber
 \label{embed16}
\end{eqnarray}
respectively, with $f(R)$ being the correction of $F(R)$ over Einstein gravity, i.e $f(R) = F(R) - R$. Moreover in the context of holographic inflation
\cite{Nojiri:2019kkp,Nojiri:2020wmh}, $c(\eta)$ and $\Lambda(\eta)$ in turn determine the corresponding holographic cut-off.\\
Coming back to the electromagnetic field action (i.e the second term in the right hand side of Eq.(\ref{full action0})), $S_{em}$ is taken as,
\begin{eqnarray}
 S_{em} = \int d^4x\sqrt{-g}\bigg[-\frac{1}{4}f_1(\eta) F_{\mu\nu}F^{\mu\nu} + f_2(\eta)F^0_{i}F^{0i} + f_3(\eta)\epsilon^{ijk}F^0_iF_{jk} 
 + f_4(\eta)\epsilon^{\mu\nu\alpha\beta}F_{\mu\nu}F_{\alpha\beta}\bigg]
 \label{em action}
\end{eqnarray}
where $F_{\mu\nu} = \partial_{\mu}A_{\nu} - \partial_{\nu}A_{\mu}$ is the field strength tensor of the electromagnetic field $A_{\mu}$.
Following the same argument as mentioned for $g^{00}$ metric component, $A^0$ component of the $A_{\mu}$ will be invariant under special 
diffeomorphism transformation.  
$f_i(\eta)$ ($i = 1,2,3,4$) are arbitrary analytic functions of $\eta$ and denote the non-minimal coupling of the electromagnetic field. 
Such time dependent coupling functions naturally break the conformal invariance in the electromagnetic sector and lead to the gauge field 
production from primordial quantum vacuum. It 
may be observed that $S_{em}$ consists of all possible electromagnetic terms (up to second order) which behave as scalar quantity under the 
spatial diffeomorphism. Moreover at later stage, we will consider 
the functions $f_i(\eta)$ in such a way that in the early universe, the couplings introduce a non-trivial correction to 
the electromagnetic field evolution, while at late times specifically at the end of inflation, $f_i(\eta)$ 
will turn out to be $f_1(\eta) = 1$, $f_2(\eta) = f_3(\eta) = f_4(\eta) = 0$ leadin to standard Maxwellian evolution.

The spacetime perturbation action $S_{sp}$, following the line of EFT, is given by,
\begin{eqnarray}
 S_{sp} = \int d^4x\sqrt{-g}&\bigg[&\frac{M_2^4(\eta)}{2}\big(\delta g^{00}\big)^2 - \frac{m_3^3(\eta)}{2}\delta K\delta g^{00} 
 - m_4^2(\eta)\big(\delta K^2 - \delta K_{\mu\nu}\delta K^{\mu\nu}\big) + \frac{\tilde{m}_4^2(\eta)}{2}R^{(3)}\delta g^{00}\nonumber\\
 &-&\bar{m}_4^2(\eta)\delta K^2 - \frac{m_5(\eta)}{2}R^{(3)}\delta K - \frac{\lambda_1(\eta)}{2}\big(R^{(3)}\big)^2 
 - \frac{\lambda_2(\eta)}{2}\nabla_{i}R^{(3)}\nabla^{i}R^{(3)}\bigg]
 \label{spacetime perturbation action}
\end{eqnarray}
where all the coefficients are allowed to vary with $\eta$. All EFT parameters are so chosen that dimensions $[m_i] = 1$ and $[\lambda_i] = 0$. 
In the above expression, $\delta K_{\mu\nu} = K_{\mu\nu} - H\sigma_{\mu\nu}$, $\delta K = K - 3H$ where 
$K_{\mu\nu}$ and $\sigma_{\mu\nu}$ are the extrinsic curvature and induced metric on a constant time hypersurface respectively. 
Moreover $\delta g^{00}$ is connected to the curvature perturbation variable $\Psi(\vec{x},\eta)$ by a non-trivial way which we will introduce later. 
The quadratic action of 
$\Psi(\vec{x},\eta)$ from Eq.(\ref{spacetime perturbation action}) contains terms like $\big(\partial\Psi(\vec{x},\eta)\big)^2$, 
$\big(\partial^2\Psi(\vec{x},\eta)\big)^2$ and $\big(\partial^3\Psi(\vec{x},\eta)\big)^2$; 
which, upon Fourier transformation behave as, 
$\big(\partial\Psi\big)^2 \sim k^2\Psi_k^2$, $\big(\partial^2\Psi\big)^2 \sim k^4\Psi_k^2$ and $\big(\partial^3\Psi\big)^2 \sim k^6\Psi_k^2$ respectively. 
Where $\Psi_k(\eta)$ is the Fourier mode with momentum $k$.
Even though they are quadratic in fluctuation, higher derivative terms will be generically suppressed by 
inflationary energy scale $H$. We, therefore, will consider the curvature perturbation action upto the quadratic order in $k$. 
Hence, we choose following condition for our subsequent discussions,\cite{Cai:2016thi,Cai:2017tku},
\begin{eqnarray}
 \bar{m}_4 = m_5 = \lambda_1 = \lambda_2 = 0
 \label{choice}
\end{eqnarray}
Correspondingly $\delta g^{00}$ is related to the metric curvature perturbation by the following way 
\cite{Cai:2016thi},
\begin{eqnarray}
 \delta g^{00} = 
 \frac{4\bigg(\frac{1}{\kappa^2} + 2m_4^2\bigg)}{a(\eta)\bigg(\frac{2H}{\kappa^2} + 4Hm_4^2 - m_3^3\bigg)}~\frac{\partial\Psi}{\partial\eta} 
 \label{delta g}
\end{eqnarray}
where ${1}/{\kappa^2} = M_\mathrm{Pl}^2$ and as mentioned earlier, 
$H$ is the Hubble parameter in cosmic time. For computational simplicity we consider inflationary phase to be nearly de-Sitter with scale factor  $a(\eta) = -{1}/{H\eta}$. Now using Eqs.(\ref{choice}) and (\ref{delta g}), the quadratic action $S_{sp}$ for the fluctuation becomes,
\begin{eqnarray}
 S[v] = \int d^4x\bigg[(v')^2 + \frac{z''}{z}v^2 - c_s^2g^{ij}\partial_iv\partial_jv\bigg] ,
 \label{curvature perturbation action}
\end{eqnarray}
where, $v(\vec{x},\eta)$ is the  Mukhanov-Sasaki variable defined as $v(\vec{x},\eta) = z(\eta)\Psi(\vec{x},\eta) = 
a(\eta)\sqrt{c_1}~\Psi(\vec{x},\eta)$. 
The quantity $c_s^2$ symbolizes the sound speed for the scalar perturbation and has the following expression,
\begin{eqnarray}
 c_s^2 = \frac{1}{c_1}\bigg(\frac{c_3'(\eta)}{a^2} - \frac{1}{\kappa^2}\bigg)
 \label{sound speed}
\end{eqnarray}
with $c_1$ and $c_3$ are expressed in terms of EFT parameters (i.e $M_2$, $m_3$, $m_4$ and $\tilde{m}_4$) as follows:
\begin{eqnarray}
 c_1(\eta) = \frac{\bigg(\frac{1}{\kappa^2} + 2m_4^2\bigg)\bigg(3m_3^6 + 8M_2^4\big[\frac{1}{\kappa^2} + 2m_4^2\big]\bigg)}
 {\bigg(\frac{2H}{\kappa^2} + 4Hm_4^2 - m_3^3\bigg)^2}~;~
 c_3(\eta) = \frac{2a(\eta)\bigg(\frac{1}{\kappa^2} + 2m_4^2\bigg)\bigg(\frac{1}{\kappa^2} + 2\tilde{m}_4^2\bigg)}
 {\bigg(\frac{2H}{\kappa^2} + 4Hm_4^2 - m_3^3\bigg)}.
 \label{c(s)}
\end{eqnarray}
 Moreover, the quadratic action for the tensor perturbation comes as,
\begin{eqnarray}
 S[v_T] = \int d^4x\bigg[(v_T')^2 + \frac{z_T''}{z_T}v_T^2 - c_T^2\big(\partial v_T\big)^2\bigg]
 \label{tensor perturbation action}
\end{eqnarray}
with $z_T^2 = a^2(\eta)\big(1 + 2\kappa^2m_4^2\big)$ and the speed of the gravitational wave ($c_T^2$) in terms of the EFT parameters 
is given by $c_T^2 = \big(1 + 2\kappa^2m_4^2\big)^{-1}$.\\
Finally $S_{int}$ refers to the quadratic interaction between the electromagnetic field and $\delta g^{00}$, which, in the context 
of EFT has the following form,
\begin{eqnarray}
 S_{int} = \int d^4x\sqrt{-g}\bigg[h(\eta)\big(\partial_{i}\delta g^{00}\big)F^{0i}\bigg]
 \label{interaction action}
\end{eqnarray}
with $h(\eta)$ being the interaction coupling strength. Such interaction of electromagnetic field ($A_{\mu}$) 
with the curvature perturbation appears naturally 
in the EFT language due to the underlying spatial diffeomorphism symmetry. However, the possible 
interaction of $A_{\mu}$ with the tensor perturbation appears in higher order action and thus we do not consider those terms for our present study. 
Along with  the quadratic terms with time dependent coupling functions $f_i(\eta)$ in Eq.(\ref{em action}), the 
term $h(\eta)\partial_{i}\delta g^{00}~F^{0i}$ (present in $S_{int}$) also contributes in breaking the conformal invariance of the electromagnetic field 
action. In terms of the Mukhanov-Sasaki variable, Eq.(\ref{interaction action}) can be written as,
\begin{eqnarray}
 S_{int} = - 2\int d^4x \frac{h(\eta)B(\eta)}{a(\eta)~z(\eta)}~\partial_i\big[v' - \frac{z'}{z}v\big]~F_{0i}~~~,~~~~~~~\mathrm{with}~~~~~
 B(\eta) = \frac{4\bigg(\frac{1}{\kappa^2} + 2m_4^2\bigg)}{\bigg(\frac{2H}{\kappa^2} + 4Hm_4^2 - m_3^3\bigg)}
 \label{interaction action modified}
\end{eqnarray}
Thus as a whole, $S_{bg}$, $S_{em}$, $S_{sp}$ and $S_{int}$ are given in Eqs.(\ref{background action}), (\ref{em action}), 
(\ref{spacetime perturbation action}) and (\ref{interaction action modified}) respectively. 
In such scenario, we aim to study inflationary magnetogenesis. In this regard let us point out some salient features of the model - 
(i) the present magnetogenesis scenario will consider the coupled evolution of electromagnetic field and the spacetime perturbation, where, 
in particular, the electromagnetic 
field gets coupled with the curvature perturbation. However the tensor perturbation evolves freely as it does not interact 
either with the electromagnetic field or with the scalar perturbation upto the quadratic order. (2) The time dependent self coupling 
of electromagnetic field denoted by the coupling strengths $f_i(\eta)$ 
and the interaction term with $\delta g^{00}$ spoil the conformal invariance of the electromagnetic field action. The coupling strengths will 
be chosen in such a way that after the inflationary epoch, the conformal invariance of $A_{\mu}$ will be restored and 
consequently the standard Maxwell's equations will be recovered. (3) Moreover as we will show in a later section 
that the well known back-reaction and strong coupling problem will be resolved in the present scenario for suitable parameter spaces.\\
In order to determine the equations of motion, let us recall that there are three independent fields in the model- the electromagnetic field $A_{\mu}$, 
the scalar and tensor Mukhanov-Sasaki variables $v(\vec{x},\eta)$ and $v_T(\vec{x},\eta)$ respectively. 
The variation of action (\ref{full action0}) with respect to the gauge field leads to the following 
equation of motion for $A_{\mu}$,
\begin{eqnarray}
 &&-\partial_{a}\big[a^4(\eta)f_1(\eta)g^{\mu a}g^{\nu b} F_{\mu\nu}\big] + 2\partial_0\big[f_2(\eta)g^{ib}~\partial_0A_i\big] 
 - 2\delta^{b}_0~\partial_j\big[f_2(\eta)g^{ij}~\partial_0A_i\big] + 2\partial_0\big[a^2(\eta)f_3(\eta)\epsilon^{0bjk}~\partial_jA_k\big]\nonumber\\ 
 &+&2\partial_j\big[a^2(\eta)f_3(\eta)\epsilon^{0ijb}~\partial_0A_j\big] - 2\delta^{b}_0~\partial_i\big[a^2(\eta)f_3(\eta)\epsilon^{0ijk}~\partial_jA_k\big] 
 + 8\partial_a\big[a^4(\eta)f_4(\eta)\epsilon^{\mu\nu\alpha\beta}~\partial_{\mu}A_{\nu}\big]\nonumber\\ 
 &+&\partial_0\big[a^2(\eta)h(\eta)g^{ib}~\partial_i\delta g^{00}\big] = 0
 \label{eom_A_1}
\end{eqnarray}
where we consider the spatially flat FRW metric ansatz as shown in Eq.(\ref{FRW metric ansatz}). 
In the Coulomb gauge ($A_0 = 0$ and $\partial^{i}A_{i} = 0$) condition, the relevant equation of motions for $A_{i}$'s are,
\begin{eqnarray}
 &f_1(\eta)&\big[A_i'' + \frac{f_1'}{f_1}A_i' - \partial_l\partial_lA_i\big] + \frac{2}{a^2(\eta)}f_2(\eta)~\big[A_i'' + \frac{f_2'}{f_2}A_i' 
 - \frac{2a'}{a}A_i'\big] + \frac{2}{a^2}f_3(\eta)\epsilon_{ijk}~\big[\frac{f_3'}{f_3} - \frac{2a'}{a}\big]~\partial_jA_k\nonumber\\ 
 &-&8f_4'(\eta)\epsilon_{ijk}~\partial_jA_k - h(\eta)~\partial_0\partial_i\big(\delta g^{00}\big) - h'(\eta)~\partial_i\big(\delta g^{00}\big) = 0
 \label{eom_A_2}
\end{eqnarray}
with a prime denoting $\frac{d}{d\eta}$.
 
The equation of motion for the scalar and tensor perturbation will take the following form,
\begin{eqnarray}
&& v''(\vec{x},\eta) - \frac{z''}{z}v - c_s^2~\partial_i\partial^{i}v = 0 \\\nonumber
&& v_T''(\vec{x},\eta) - \frac{z_T''}{z_T}v_T - c_T^2~\partial_i\partial^{i}v_T = 0~~.
\label{eom_tp}
\end{eqnarray}
where in the above derivation Coulomb gauge condition is used. 
with recall, $z(\eta) = a(\eta)\sqrt{c_1(\eta)}$ and $c_s^2$ is given in Eq.(\ref{sound speed}). 
It may be observed that the scalar perturbation evolves freely, but the electromagnetic field evolution does depend on the scalar perturbation 
evolution through $h(\eta)$ as evident from Eq.(\ref{eom_A_2}). This is due to  Coulomb gauge condition. 
The solution of these equations require a certain ansatz of the background spacetime scale factor $a(\eta)$ and the 
coupling functions $f_i(\eta)$, $h(\eta)$. For the background spacetime, we consider a de-Sitter inflationary 
spacetime, in which case, the scale factor is given by,
\begin{eqnarray}
 a(\eta) = -\frac{1}{H\eta}~~~,
 \label{scale factor}
\end{eqnarray}
with $H$ being is very slowly varying function and represents the Hubble parameter in cosmic time. 
The scale factor of Eq.(\ref{scale factor}) immediately leads to the Hubble parameter in conformal time as 
$\mathcal{H} = \frac{1}{a}\frac{da}{d\eta} = -1/\eta$. 
Furthermore, we introduce the inflationary e-folding number as $N(\eta) = \ln{\big(a(\eta)\big)}$, which is counted from the beginning of inflation where 
$N = 0$ and we consider the beginning of inflation to be the instant when the CMB scale mode crosses the horizon.
 
Here we assume the EFT parameters are non-zero (i.e all the electromagnetic terms present in the 
EFT action (\ref{em action}) have been taken into account) and have the power law forms of the scale factor, in particular, 
\begin{eqnarray}
 f_1(\eta)=\left\{\begin{array}{cc}\bigg(\frac{a(\eta)}{a_f(\eta_f)}\bigg)^2 & ~~~\eta \leq \eta_f \\
1 &~~~ \eta \geq \eta_f 
\end{array} \right.
 \label{f1}
 \end{eqnarray}
 and
 \begin{eqnarray}
 f_2(\eta)=\bigg(\frac{a_i(\eta_i)}{a(\eta)}\bigg)^m~~;~~f_3(\eta) = \bigg(\frac{a_i(\eta_i)}{a(\eta)}\bigg)^r ~~;~~
 f_4(\eta)=\bigg(\frac{a_i(\eta_i)}{a(\eta)}\bigg)^s~~;~~h(\eta) = h_0\bigg(\frac{a_i(\eta_i)}{a(\eta)}\bigg)^2
 \label{coupling functions}
\end{eqnarray}
where $a_i$ and $a_f$ are the scale factors at the beginning $\eta_i$ and at the end of inflation $\eta_f$ respectively. 
Moreover $h_0$ is constant having mass dimension unity. The exponents $m$, $r$, $s$ are positive numbers 
and can act as model parameters in the present context. Later, we will show that $m = r = s = 2$ leads to a scale invariant 
magnetic power spectrum in the superhorizon limit. It is evident 
from the above expressions that $f_1(\eta)$ starts from $e^{-2N_f}$ and monotonically increases till $\eta = \eta_f$, 
while the other coupling functions monotonically decreases with time during the inflationary stage. 
However in the post-inflationary phase, $f_1(\eta)$ becomes unity and $f_2(\eta) \approx f_3(\eta) \approx f_4(\eta) \approx h(\eta) \approx 0$, 
which in turn recovers the standard Maxwell's equations at late time.

Before embarking on our study with the aforementioned forms of the EFT parameters, let us contextualise our discussions by giving few explicit model examples with elaborate discussions given in appendix-\ref{appendix}.\\
 	\textbf{Generalized Ratra Model}: In this model the conformal invariance is broken by generic scalar function as a gauge kinetic function as 
\begin{eqnarray}
		S_{m1} = \int d^4x\sqrt{-g}\left[f(\phi, R, {\cal G})F_{\mu\nu}F^{\mu\nu}\right],
		\label{new action1}
		\end{eqnarray}
	where $(\phi, R, {\cal G})$ are scalar field under consideration (generally the inflaton), the background 
	Ricci scalar and the Gauss-Bonnet scalar respectively. The scenario with $f(\phi,R,{\cal G}) = f(\phi)$ is well known Ratra model whch 
	has been studied extensively (without or with reheating phase) in \cite{Demozzi:2009fu,Haque:2020bip,Kobayashi:2019uqs,Ratra:1991bn}; 
	moreover the case $f(\phi,R,\mathcal{G}) = f(R,\mathcal{G})$ has also been explored (without or with reheating phase) in \cite{Bamba:2020qdj}. 
	Comparing the action (\ref{new action1}) with the EFT action (\ref{em action}), the associated EFT parameters can be mapped as:
		\begin{eqnarray}
		f_1(\eta) = f\left(\phi(\eta),R(\eta),{\cal G}(\eta)\right)~~~~~~,~~~~~f_2(\eta) = f_3(\eta) = f_4(\eta) = 0 .
		\label{correspondence1} .	\end{eqnarray}
where $\phi(\eta)$ can be obtained from the background evolution of the scalar field over FRW spacetime, 
$R(\eta) = \frac{6}{a^2}\left(\mathcal{H}' + \mathcal{H}^2\right)$ and $\mathcal{G}(\eta) = \frac{24}{a^4}\mathcal{H}^2\mathcal{H}'$, 
with $\mathcal{H}$ represents the conformal Hubble parameter. One gets a further generalized scenario by 
adding parity violating term in the above Lagrangian, in particular,
\begin{eqnarray}
 S_{g} = S_{m1} + S_{pv}
 \label{new action5}
\end{eqnarray}
with
\begin{eqnarray}
 S_{pv} = \int d^4x\sqrt{-g}\bigg[\frac{\alpha}{M^2}\epsilon^{\mu\nu\alpha\beta}RF_{\mu\nu}F_{\alpha\beta} 
 + \frac{\beta}{M^2}\epsilon^{\mu\alpha\beta\delta}R_{\mu\nu}F_{\alpha}^{\nu}F_{\beta\delta} 
 + \frac{\gamma}{M^2}\epsilon^{\mu\nu\alpha\beta}R_{\mu\nu}^{~~\rho\sigma}F_{\rho\sigma}F_{\alpha\beta}\bigg]
\end{eqnarray}
with $\alpha$, $\beta$ and $\gamma$ are model parameters. The background spacetime immediately leads to the Ricci scalar and the non-zero 
components of Ricci tensor, Riemann tensor as,
\begin{eqnarray}
 R&=&6\frac{a''}{a^3}~~~~~~~,~~~~~~~~R_{00} = -3\left(\frac{a''}{a} - \frac{a'^2}{a^2}\right)~~~~~~~,~~~~~~~
 R_{ij} = \left(\frac{a''}{a} + \frac{a'^2}{a^2}\right)\nonumber\\
 R_{0i}^{~~0j}&=&\left(\frac{a'^2}{a^4} - \frac{a''}{a^3}\right)\delta_{i}^{j}~~~~~~~,~~~~~~~
 R_{ij}^{~~kl} = \frac{a'^2}{a^4}\left(\delta_{i}^{l}\delta_{j}^{k} - \delta_{i}^{k}\delta_{j}^{l}\right)~~.
\end{eqnarray}
Consequently, the action $S_{pv}$ turns out to be,
\begin{eqnarray}
 S_{pv}&=&\int d^4x\sqrt{-g}\bigg[6\frac{\alpha}{M^2}\epsilon^{\mu\nu\alpha\beta}\left(\frac{a''}{a^3}\right)F_{\mu\nu}F_{\alpha\beta}\nonumber\\ 
 &+&\frac{1}{M^2}\epsilon^{ijk}\bigg\{2\beta\left(aa'' + a'^2\right)-3\beta\left(\frac{a''}{a} - \frac{a'^2}{a^2}\right) 
 - \gamma\left(\frac{a''}{a} + \frac{a'^2}{a^2}\right)\bigg\}F_{i}^{0}F_{jk}\bigg]~~.
 \label{new action5 modified}
\end{eqnarray}
Thereby comparing the action (\ref{new action5}) with the EFT action of Eq.(\ref{em action}), we argue that the action 
$S_{m5}$ can be embedded within the EFT action of EM field, provided the EFT parameters have the following forms,
\begin{eqnarray}
 f_1(\eta)&=&f\left(\phi(\eta),R(\eta),{\cal G}(\eta)\right)~~~~~~,~~~~~~f_2(\eta) = 0\nonumber\\
 f_3(\eta)&=&\frac{1}{M^2}\bigg\{2\beta\left(aa'' + a'^2\right)-3\beta\left(\frac{a''}{a} - \frac{a'^2}{a^2}\right) 
 - \gamma\left(\frac{a''}{a} + \frac{a'^2}{a^2}\right)\bigg\}\nonumber\\
 f_4(\eta)&=&6\frac{\alpha}{M^2}\left(\frac{a''}{a^3}\right)
 \label{correspondence5}
\end{eqnarray}
respectively. By this way, one may construct more general magnetogenesis models from the EFT formalism with suitable choices of $f_i(\eta)$. This 
requires a complete scan of the EFT parameters for which the EFT action can generate sufficient magnetic strength and at the same time be self consistent, 
which we expect to study in future. However the present work and the following discussions are based on the 
EFT parameters chosen in Eqs. \eqref{f1} and \eqref{coupling functions} respectively.

\section{Energy density and power spectra for electromagnetic and metric perturbation fields}\label{sec_power spectra}

In the present section, we will calculate the power spectra for both the electromagnetic and metric perturbation fields. In regard to the electromagnetic 
field, it may be mentioned that the electric and the magnetic fields are frame dependent. In the present context, the electric and magnetic fields 
are referred with respect to the comoving observer, in which case the proper time becomes identical with the cosmic time or equivalently 
the four velocity components of a comoving observer are given by $u^{\mu} = \big(1/a(\eta), 0, 0, 0\big)$. 
For the purpose of the computation of the power spectrum; first we need to know the energy density 
for the respective fields and, secondly, the vacuum state associated with the field in the background inflationary evolution. 
Thereby, from the action (\ref{full action0}), we first determine the energy-momentum tensor associated with the electromagnetic field,
\begin{eqnarray}
 T_{ab}&=&
-\frac{1}{4}f_1(\eta)~\big[g_{ab}F^2 - 4g^{\mu\alpha}F_{\mu a}F_{\alpha b}\big] 
 + f_2(\eta)~\big[\frac{1}{a^4}g_{ab}g^{ij}F_{0i}F_{0j} - \frac{2}{a^4}F_{0a}F_{0b} - \frac{4}{a^2}~\delta^0_ag^{ij}F_{0i}F_{bj}\big]\nonumber\\ 
 &+&\frac{2}{a^4}~f_3(\eta)\epsilon_{ijk}~\delta^0_a\ F_{b i}F_{jk} 
 + h(\eta)~\big[g_{ab}g^{0\mu}g^{i\nu}F_{\mu\nu}~\partial_i\big(\delta g^{00}\big) - 2\delta^0_ag^{i\nu}F_{b\nu}~\partial_i\big(\delta g^{00}\big) 
 - 2g^{0\mu} F_{\mu b}~\partial_a\big(\delta g^{00}\big)\big]\nonumber\\
 \label{em_tensor}
\end{eqnarray}
The energy density of the EM field in the background FRW spacetime is given by $T^0_0 = -\frac{1}{a^2}T_{00}$ and thus 
the above expression of $T_{\alpha\beta}$ immediately leads to the following form of $T^0_0$:
\begin{eqnarray}
 T^0_0 &=& f_1(\eta)~\big[-\frac{1}{2a^4}(A_i')^2 - \frac{1}{4a^4}F_{ij}F_{ij}\big] + f_2(\eta)~\big[-\frac{3}{a^6}(A_i')^2\big] \\\nonumber
 &+& f_3(\eta)~\big[-\frac{4}{a^6}\epsilon_{ijk}A_i'~\partial_jA_k\big] + h(\eta)~\big[\frac{1}{a^4}A_i'~\partial_i\big(\delta g^{00}\big)\big]
 \label{energy_density_A}
\end{eqnarray}
where we use the Coulomb gauge condition. Having determined $T_0^0$, we express the total electromagnetic energy density in terms of  electric, 
magnetic and interaction energy density as,
\begin{eqnarray}
&& \rho^{em}_{total} = \rho(\vec{E}) + \rho(\vec{B}) + \rho_{int}(\vec{E},\vec{B})+ \rho_{int}(\vec{E},\delta g^{00}) \\
 &&\rho(\vec{E}) =\langle 0\big|f_1(\eta)~\big[-\frac{1}{2a^4}(A_i')^2\big] + f_2(\eta)~\big[-\frac{3}{a^6}(A_i')^2\big]\big|0 \rangle~~;~~
 \rho(\vec{B}) = \langle 0\big|f_1(\eta)~\big[-\frac{1}{4a^4}F_{ij}F_{ij}\big]\big|0 \rangle\nonumber\\
 &&\rho_{int}(\vec{E},\vec{B})=\langle 0\big|f_3(\eta)~\big[-\frac{4}{a^6}\epsilon_{ijk}A_i'~\partial_jA_k\big]\big|0 \rangle~~;~~
 \rho_{int}(\vec{E},\delta g^{00}) = \langle 0\big|h(\eta)\big[\frac{1}{a^4}A_i'~\partial_i\big(\delta g^{00}\big)\big]\big|0 \rangle
 \label{part_energy density A}
\end{eqnarray}
Here $|0\rangle$ is the quantum vacuum defined at the distant past which is Bunch-Davies state. For quantization we 
promote $A_i(\eta,\vec{x})$ and $v(\vec{x},\eta)$ to hermitian operators $\hat{A}_i(\vec{x},\eta)$ and $\hat{v}(\vec{x},\eta)$ and expanding 
them in a Fourier basis as follows,
\begin{eqnarray}
 \hat{A}_i(\vec{x},\eta) &=& \int \frac{d^3k}{(2\pi)^3} \sum_{p=+,-}\epsilon^{(p)}_i~\bigg[\hat{b}_p(\vec{k})A_p(k,\eta)e^{i\vec{k}.\vec{x}} 
 + \hat{b}_p^{\dagger}(\vec{k})A_p^{*}(k,\eta)e^{-i\vec{k}.\vec{x}}\bigg]
 \label{mode_expansion_A} \\
 \hat{v}(\vec{x},\eta) &=& \int \frac{d^3k}{(2\pi)^3}~\bigg[\hat{c}(\vec{k})\tilde{v}(k,\eta)e^{i\vec{k}.\vec{x}} 
 + \hat{c}^{\dagger}(\vec{k})\tilde{v}^{*}(k,\eta)e^{-i\vec{k}.\vec{x}}\bigg]
 \label{mode_expansion_sp}.
\end{eqnarray}
 $\epsilon^{(p)}_i$ is the polarization vector with $p$ being polarization index $p =\pm$.
Here we consider the polarization vectors 
in helicity basis, in which case $\epsilon^{+}_i = {1/\sqrt{2}}(1, i, 0)$ and $\epsilon^{-}_i = {1/\sqrt{2}}(1, -i, 0)$. 
The Coulomb gauge indicates that the propagating direction or the momentum of the electromagnetic wave is perpendicular to 
its polarization vector $\epsilon^{\pm}_i$ i.e $k^{i}\epsilon^{\pm}_i = 0$. Consequently, the polarization vectors further 
satisfies the following relation, $\epsilon_{ijk}k_{j}~\epsilon^{\pm}_k = \mp ik \epsilon^{\pm}_i$. 
Moreover, $\hat{b}_p(\vec{k})$, $\hat{b}_p^{\dagger}(\vec{k})$ and $\hat{c}(\vec{k})$, $\hat{c}^{\dagger}(\vec{k})$ 
are the annihilation, creation operators for the respective fields defined in the distant past with respect to Bunch-Davies vacuum state $|0\rangle$, i.e 
$\hat{b}_r(\vec{k})|0\rangle = 0$ and $\hat{c}(\vec{k})|0\rangle = 0,~ \forall ~\vec{k}$. 
Such creation and annihilation operators follow the quantization rule, as
\begin{eqnarray}
 \big[\hat{b}_p(\vec{k}),\hat{b}_q^{\dagger}(\vec{k}')\big] = \delta_{pq}~\delta(\vec{k} - \vec{k}')~~,~~ 
 \big[\hat{c}(\vec{k}),\hat{c}^{\dagger}(\vec{k}')\big] = \delta(\vec{k} - \vec{k}')
 \label{commutation}
\end{eqnarray}
and all the other commutators are zero. With these mode decomposition of $A_i(\vec{x},\eta)$ and $v(\vec{x},\eta)$ , the individual 
component of the electromagnetic energy densities in the Bunch-Device vacuum state turn out to be,
\begin{eqnarray}
&&\rho(\vec{E})=\bigg\{\frac{f_1(\eta)}{a^4} + \frac{6f_2(\eta)}{a^6}\bigg\}~\sum_{p=+,-}\int\frac{k^2}{2\pi^2}\big|A_p'(k,\eta)\big|^2~dk\nonumber\\
&& \rho(\vec{B})=\frac{f_1(\eta)}{a^4}~\sum_{p=+,-}\int\frac{k^4}{2\pi^2}\big|A_p(k,\eta)\big|^2~dk\nonumber\\
&& \rho_{int}(\vec{E},\vec{B})=\frac{4f_3(\eta)}{a^6}~\sum_{p=+,-}\int\frac{k^3}{2\pi^2}\big|A_p(k,\eta)A_p'(k,\eta)\big|~dk\nonumber\\
&& \rho_{int}(\vec{E},\delta g^{00})=0
 \label{expectation_energy A}
\end{eqnarray}
The vacuum expectation for the interaction energy between electric field and the scalar perturbation is zero because of the fact that 
the creation/annihilation operators of the scalar perturbation i.e $\hat{c}^{\dagger}(\vec{k}),\hat{c}(\vec{k})$ commute with that 
of the electromagnetic field. As can be observed that the energy $\rho_{int}(\vec{E},\vec{B})$ 
dilutes much faster than the other energy component. Hence, we will ignore this term in our subsequent discussion. 
The power spectra, defined as the energy density associated to a logarithmic interval of $k$, of the electric and magnetic fields follow
\begin{eqnarray}
 P^{(E)}(k,\eta)&=&\bigg\{\frac{f_1(\eta)}{a^4} + \frac{6f_2(\eta)}{a^6}\bigg\}\sum_{p=+,-}\frac{k^3}{2\pi^2}\big|A_p'(k,\eta)\big|^2\nonumber\\
 P^{(B)}(k,\eta)&=&\frac{f_1(\eta)}{a^4}\sum_{p=+,-}\frac{k^5}{2\pi^2}\big|A_p(k,\eta)\big|^2~~~.
 \label{power spectra em}
\end{eqnarray}
Furthermore, the scalar and tensor power spectra are given by,
\begin{eqnarray}
 P_s(k,\eta) = \frac{k^3}{2\pi^2}\bigg|\frac{\tilde{v}(k,\eta)}{z(\eta)}\bigg|^2~~~,~~~
 P_T(k,\eta) = \frac{k^3}{2\pi^2}\bigg|\frac{\tilde{v}_T(k,\eta)}{z_T(\eta)}\bigg|^2~~~.
 \label{power spectra metric perturbation}
\end{eqnarray}
As the effective theory Lagrangian has parity violating operators, the gauge field components in helicity basis $(A_+, A_-)$ evolve differently. 
Hence, the quantity which measures this is related to helicity density which is defined as $\rho_{h}(\eta,k) = \langle0|A_{i}B^{i}|0\rangle$ 
in Bunch-Davies vacuum. The expression take the following well known form,
\begin{eqnarray}
\rho_{h}(k,\eta) = \frac{1}{2\pi^2}\int \frac{dk}{k}\bigg(\frac{k^4}{a^3}\bigg)\bigg(\big|A_{+}(\eta,k)\big|^2 - \big|A_{-}(\eta,k)\big|^2\bigg)~~.
\label{helicity1}
\end{eqnarray}
Following the definition of electric and magnetic power spectrum, we can find out the helicity spectrum as
\bea
P^{(h)} = \frac{\partial\rho_{h}}{\partial\ln{k}} = 
\bigg(\frac{k^4}{2\pi^2 a^3}\bigg)\bigg\{\big|A_{+}(k,\eta)\big|^2 - \big|A_{-}(k,\eta)\big|^2\bigg\}~~.
\label{helicity2}
\eea
We have all the necessary expressions of the electromagnetic spectrum which indeed depend on the evolution of the electromagnetic mode function. 
Thereby the explicit $k$ and $\eta$ dependence of the power spectra demands the solution of the mode function. This will be discussed in the next section.

\section{Solving for the electromagnetic mode function and metric perturbation variables}\label{sec_solution_inflation}
\begin{figure}[t]
\begin{center}
 \centering
 \includegraphics[width=3.5in,height=2.5in]{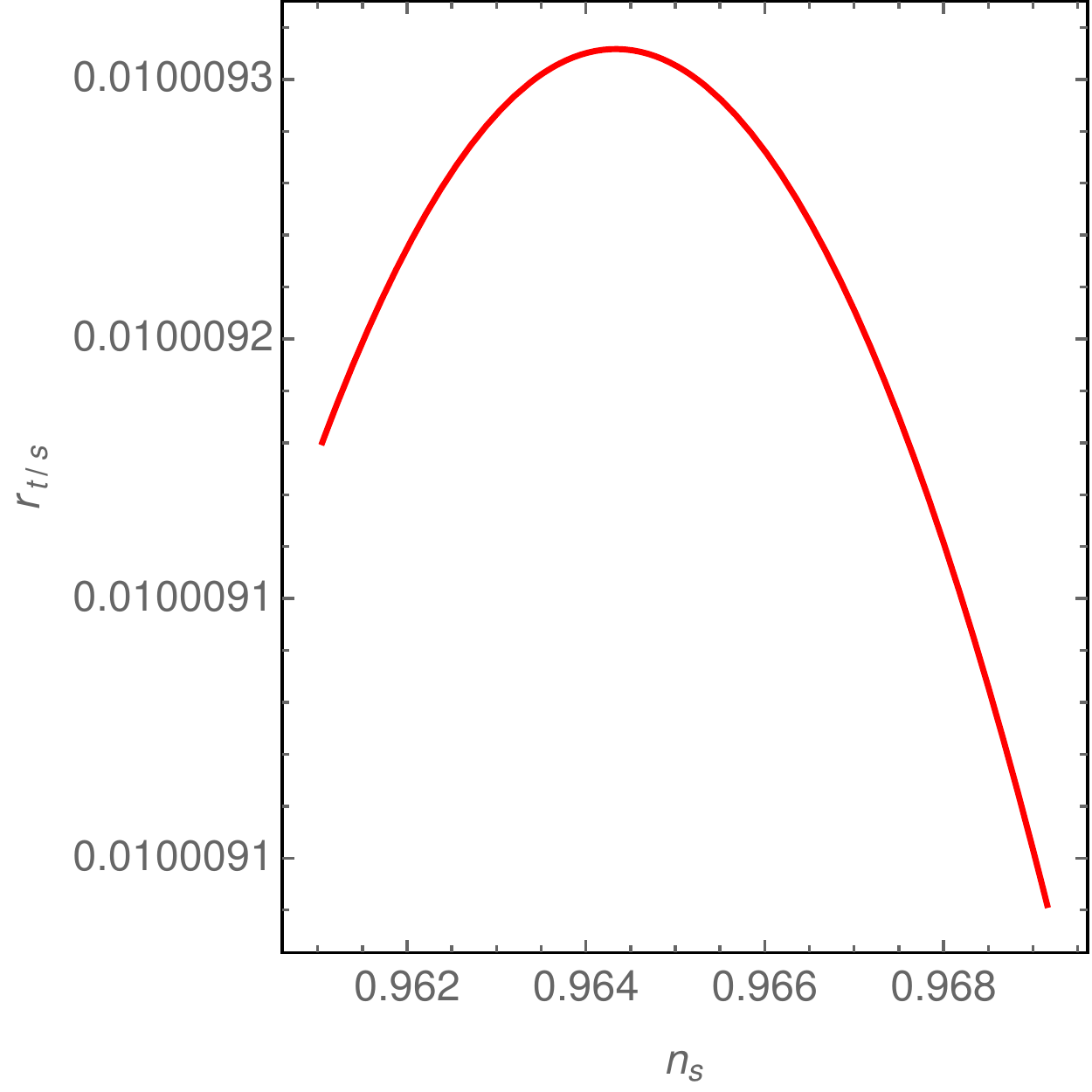}
 \caption{Parametric plot of $n_s$ vs $r_\mathrm{t/s}$ for $0.0155 \leq \gamma \leq 0.0196$.}
 \label{plot_observable}
\end{center}
\end{figure}

The evolution of the vector potential in terms of the conformal time is given in Eq.(\ref{eom_A_2}), which, in Fourier space, can be recast as,
\begin{eqnarray}
 &f_1(\eta)&\bigg[A_{\pm}'' + \frac{f_1'}{f_1}A_{\pm}' + k^2A_{\pm}\bigg] 
 + \frac{2}{a^2}f_2(\eta)~\bigg[A_{\pm}'' - A_{\pm}'\big(\frac{2a'}{a} - \frac{f_2'}{f_2}\big)\bigg] 
 \mp \frac{2}{a^2}f_3(\eta)kA_{\pm}~\bigg[\frac{2a'}{a} - \frac{f_3'}{f_3}\bigg] \mp 8f_4'(\eta)kA_{\pm}\nonumber\\ 
 &\mp&\frac{\sqrt{2}B(\eta)}{az}h'(\eta)k~\bigg[\tilde{v}' - \frac{z'}{z}\tilde{v}\bigg] 
 \mp \frac{\sqrt{2}B(\eta)}{az}h(\eta)k~\bigg[\big(\tilde{v}'' - \frac{z''}{z}\tilde{v}\big) 
 + \big(\tilde{v}' - \frac{z'}{z}\tilde{v}\big)\big(\frac{B'}{B} - \frac{a'}{a} - \frac{2z'}{z}\big)\bigg] = 0
 \label{eom_A_FM}
\end{eqnarray}
where $B(\eta)$ is shown in Eq.(\ref{interaction action modified}) and $k = \big|\vec{k}_{(em)}\big| = \big|\vec{k}_{(sp)}\big|$ 
in the above expression, with $\big|\vec{k}_{(em)}\big|$ and 
$\big|\vec{k}_{(sp)}\big|$ being the momentum modulus of the electromagnetic and 
the scalar perturbation field respectively. Is is evident 
from Eq.(\ref{eom_A_FM}) that the dynamical equation of $A_{+}(k,\eta)$ and $A_{-}(k,\eta)$ differ from each other due to the presence 
of the couplings $f_3(\eta)$, $f_4(\eta)$ and $h(\eta)$ in the electromagnetic action. Similarly the 
scalar and tensor Mukhanov-Sasaki equation in Fourier space can be obtained from  (\ref{eom_tp}) as
given by,
\begin{eqnarray}
 \tilde{v}''(k,\eta) + \bigg(c_s^2(\eta)k^2 - \frac{z''}{z}\bigg)\tilde{v}(k,\eta) = 0
 \label{eom_sp_FM}\\
  \tilde{v}_T''(k,\eta) + \bigg(c_T^2(\eta)k^2 - \frac{z_T''}{z_T}\bigg)\tilde{v}_T(k,\eta) = 0
 \label{eom_tp_FM}
\end{eqnarray}
respectively. Recall, $c_s^2 = \frac{1}{c_1}\big(\frac{c_3'}{a^2} - \frac{1}{\kappa^2}\big)$, $c_T^2 = \frac{1}{1 + 2\kappa^2m_4^2}$, 
$z^2(\eta) = a^2c_1(\eta)$ and $z_T^2(\eta) = a^2\big(1 + 2\kappa^2m_4^2\big)$, where $c_1$ and $c_3$ are formed by the EFT coefficients 
(i.e $M_2$, $m_3$, $m_4$ and $\tilde{m}_4$) as shown earlier in Eq.(\ref{c(s)}).\\
With the above equations of motion, we first solve the scalar, tensor perturbation equations and then by using the solution of 
$\tilde{v}(k,\eta)$, we move on to determine the electromagnetic mode function from Eq.(\ref{eom_A_FM}). However for the purpose of solving 
the scalar and tensor perturbation variables, one needs an explicit form of $c_s^2$, $c_T^2$ and $z^2$ in terms of conformal time. In the present context, 
we consider, $
 c_s^2 = 1, c_T^2 = 1,
 $
which in turn put certain conditions on the EFT coefficients. However, at this point let us point out that  such conditions, is supported by 
various modified gravity theories such as canonical scalar-tensor theory, F(R) higher curvature 
theory or more generally $F(\phi,R)$ theory etc. Moreover the unit gravitational wave speed i.e $c_T^2 = 1$ is also compatible with the recent 
gravitational wave observation by LIGO. With $c_s^2 = 1$, the scalar power 
spectrum becomes $exactly$ scale invariant if $z(\eta)$ behaves as $z(\eta) \propto \eta^{-1}$, 
which is not consistent with the Planck observations. 
Thereby in order to make the theoretical predictions compatible with the Planck results, 
we consider, $z(\eta) = \kappa^{\gamma}(-{1}/{\eta})^{1+\gamma}$, with $\gamma$, a small parameter, makes the scalar power 
spectrum compatible with the latest PLANCK observations. 
With those forms, the EFT coefficients will take the following form,
\begin{eqnarray}
 \frac{1}{c_1(\eta)}\bigg(\frac{c_3'(\eta)}{a^2} - \frac{1}{\kappa^2}\bigg) = 1~~,~~
 \frac{1}{1 + 2\big(\kappa m_4(\eta)\big)^2} = 1~~,~~
 c_1(\eta) = H^2\kappa^{2\gamma}\bigg(\frac{-1}{\eta}\bigg)^{2\gamma}
 \label{EFT constraints}
\end{eqnarray}
Correspondingly, one of the possible choices on $m_3$, $m_4$, $M_2$ and $\tilde{m}_4$ by which Eq.(\ref{EFT constraints}) is satisfied are 
given by,
\begin{eqnarray}
 m_3 = 0~~,~~m_4 = 0~~,~~M_2(\eta) = \frac{1}{2^{1/4}}H\kappa^{\gamma/2}\bigg(\frac{-1}{\eta}\bigg)^{\gamma/2} \nonumber\\ 4\tilde{m}_4\frac{d\tilde{m}_4}{d\eta} + 2\tilde{m}_4^2\bigg(\frac{-1}{\eta}\bigg) = H^2\kappa^{2\gamma}\bigg(\frac{-1}{\eta}\bigg)^{1+2\gamma},
 \label{EFT constraints modified2}
\end{eqnarray}
where it may be observed that $\tilde{m}_4$ obeys a first order differential Eq.(\ref{EFT constraints modified2}).
With these choice of EFT parameters, the scalar Mukhanov-Sasaki equation takes the following form,
\begin{eqnarray}
 \tilde{v}''(k,\eta) + \bigg(k^2 - \frac{2\big(1 + 3\gamma/2\big)}{\eta^2}\bigg)\tilde{v}(k,\eta) = 0,
 \label{eom_sp_FM modified}
\end{eqnarray}
which can be exactly solved as,
\begin{eqnarray}
 \tilde{v}(k,\eta) = \sqrt{-k\eta}\bigg[D_1~J_{\nu}(-k\eta) + D_2~J_{-\nu}(-k\eta)\bigg],
 \label{solution_sp_FM}
\end{eqnarray}
where $\nu = \sqrt{\frac{9}{4} + 3\gamma}$, $J_{\nu}$ is the Bessel function of the first kind. $D_1$, $D_2$ are two integration constants 
which are determined by setting Bunch-Davies initial condition at in the infinite past $|k\eta| >> 1$,
\begin{eqnarray}
 D_1 = \frac{1}{2}\sqrt{\frac{\pi}{k}}~\frac{e^{-i\frac{\pi}{2}(\nu + \frac{1}{2})}}{\cos{\big[\pi(\nu + 1/2)\big]}}~~~~~~~~~~~~,~~~~~~~~~~~
 D_2 = \frac{1}{2}\sqrt{\frac{\pi}{k}}~\frac{e^{i\frac{\pi}{2}(\nu + \frac{3}{2})}}{\cos{\big[\pi(\nu + 1/2)\big]}},
 \label{integration constants}
\end{eqnarray}
and, consequently, the final solution of the scalar Mukhanov-Sasaki variable becomes
\begin{eqnarray}
 \tilde{v}(k,\eta) = \frac{\sqrt{-k\eta}}{2}~
 \bigg\{\sqrt{\frac{\pi}{k}}~\frac{e^{-i\frac{\pi}{2}(\nu + \frac{1}{2})}}{\cos{\big[\pi(\nu + 1/2)\big]}}~J_{\nu}(-k\eta) 
 + \sqrt{\frac{\pi}{k}}~\frac{e^{i\frac{\pi}{2}(\nu + \frac{3}{2})}}{\cos{\big[\pi(\nu + 1/2)\big]}}~J_{-\nu}(-k\eta)\bigg\}~~~.
 \label{final solution_sp_FM}
\end{eqnarray}
Finally, in the superhorizon limit, when the modes are going outside the Hubble radius i.e $k < \frac{1}{\mathcal{H}}$ 
(recall $\mathcal{H}$ is the Hubble parameter), the $\tilde{v}(k,\eta)$ can be expressed as 
\begin{eqnarray}
 \lim_{|k\eta| \ll 1} \tilde{v}(k,\eta) = 
 \bigg\{\frac{D_1}{2^{\nu}\Gamma(\nu + 1)}~\big(-k\eta\big)^{\nu + \frac{1}{2}} + \frac{D_2}{2^{-\nu}\Gamma(-\nu + 1)}
 ~\big(-k\eta\big)^{-\nu + \frac{1}{2}}\bigg\},
 \label{superhorizon form}
\end{eqnarray}
where we have used the power law expansion of the Bessel function given by 
$\lim_{|k\eta| \ll 1} J_{\nu}(-k\eta) = \frac{1}{2^{\nu}\Gamma(\nu + 1)}(-k\eta)^{\nu}$, and similarly for 
$J_{-\nu}(-k\eta)$. However due to the fact $\nu > 1/2$, the second term containing $\big(-k\eta\big)^{-\nu + \frac{1}{2}}$ in the 
above expression becomes dominant over the other one and thus $\tilde{v}(k,\eta)$ effectively behaves as 
$\tilde{v}(k,\eta) \sim \big(-k\eta\big)^{-\nu + \frac{1}{2}}$ in the superhorizon limit. Consequently the scalar power spectrum 
is determined as,
\begin{eqnarray}
 P_{\Psi}(k,\eta) = \frac{k^3}{2\pi^2}\bigg|\frac{\tilde{v}(k,\eta)}{z(\eta)}\bigg|^2 
 \label{power spectrum sp}
\end{eqnarray}
with $\tilde{v}(k,\eta)$ is given in Eq.(\ref{final solution_sp_FM}). In view of Eq.(\ref{superhorizon form}), the scalar power spectrum 
goes as $P_{\Psi} \sim \big(k|\eta|\big)^{3-2\nu}$ in the superhorizon scale, which in turn provides the 
scalar spectral index as $n_s = 4 - 2\nu$. With the help of the solution of $\tilde{v}(k,\eta)$, we will solve the electromagnetic mode function 
from Eq.(\ref{eom_A_FM}). However before going to the electromagnetic mode function, for the sake of completeness let us 
determine the tensor power spectrum which can be obtained from the solution the perturbation Eq.(\ref{eom_tp_FM}),
\begin{eqnarray}
 \tilde{v}_T(k,\eta) = \frac{\sqrt{-\pi\eta}}{2}H^{(2)}\bigg[\frac{3}{2},-k\eta\bigg]
 \label{final solution_tp_FM}
\end{eqnarray}
where $H^{(2)}\big[\frac{3}{2},-k\eta\big]$ symbolizes the Hankel function of second kind with order $3/2$. In the infinite 
past the above solution matches with the Bunch-Davies initial 
condition, $\lim_{|k\eta| \gg 1} \tilde{v}_T(k,\eta) = \frac{1}{\sqrt{2k}}e^{-ik\eta}$. Thus the tensor power spectrum 
comes with the following expression,
\begin{eqnarray}
 P_{T}(k,\eta) = 2\times\frac{k^3}{2\pi^2}\bigg|\frac{\tilde{v}_T(k,\eta)}{z_T(\eta)}\bigg|^2 
 = \frac{k^3}{4\pi^2}\bigg|\frac{\sqrt{-\pi\eta}}{z_T(\eta)}~H^{(2)}\big[\frac{3}{2}, -k\eta\big]\bigg|^2~~~.
 \label{power spectrum tp}
\end{eqnarray}
The multiplicative factor $2$ arises due to the two polarization waves in the gravity wave.
Now we can confront the model at hand with the latest 
Planck observational data \cite{Akrami:2018odb}, so we shall 
calculate the spectral index of the primordial curvature perturbations $n_s$ and the tensor-to-scalar ratio $r_\mathrm{t/s}$, which are defined as follows,
\begin{align}
n_s = 1 + \left. \frac{\partial \ln{P_{\Psi}}}{\partial
\ln{k}}\right|_{*} \, , \quad
r_\mathrm{t/s} = \left. \frac{P_T(k,\eta)}{P_{\Psi}(k,\eta)}\right|_{*}\, ,
\label{obs1}
\end{align}
where the suffix $*$ corresponds to the horizon crossing instant of the CMB scale (afterwards in the paper, a suffix $*$ with a quantity 
represents that quantity at the instant when the CMB scale crosses the horizon). Eqs.(\ref{power spectrum sp}) and 
(\ref{power spectrum tp}) immediately lead to the explicit forms of $n_s$ and $r_\mathrm{t/s}$ as follows,
\begin{eqnarray}
n_s&=&4 - 3\sqrt{1 + \frac{4\gamma}{3}}\nonumber\\
r_\mathrm{t/s}&=&\bigg|\frac{z(\eta_*)}{z_T(\eta_*)}(\sqrt{-\pi\eta_*})\tilde{v}(k,\eta_*)~H^{(2)}\big[\frac{3}{2}, -k\eta_*\big]\bigg|^2
\label{obs2}
\end{eqnarray}
with $\tilde{v}(k,\eta)$ is given in Eq.(\ref{final solution_sp_FM}). 
It may be noticed that $n_s$ and $r_\mathrm{t/s}$ depend on the dimensionless parameter $\gamma$. Here we would like to mention that in the context of EFT, the 
spectral index can be expressed as \cite{Cheung:2007st},
\begin{eqnarray}
 n_s = 1 + 4\frac{\dot{H}_*}{H_*^2} - \frac{\ddot{H}_*}{H_*\dot{H}_*}
 \label{ns EFT}
\end{eqnarray}
Comparing Eqs.(\ref{obs2}) and (\ref{ns EFT}) immediately connects 
$\gamma$ with $\dot{H}_*$ and $\ddot{H}_*$, as given by the following relation
\begin{eqnarray}
 \gamma = -\frac{2\dot{H}_*}{H_*^2} + \frac{\ddot{H}_*}{2H_*\dot{H}_*}~~.
 \label{gamma EFT comparison}
\end{eqnarray}
Having obtained Eq.(\ref{obs2}), we can now directly confront the spectral index and 
the tensor-to-scalar ratio with the Planck 2018 constraints \cite{Akrami:2018odb}, which constrain the observational indices as follows,
\begin{equation}
\label{planckconstraints}
n_s = 0.9649 \pm 0.0042\, , \quad r_\mathrm{t/s} < 0.064\, .
\end{equation}
In the present context, the theoretical expectations of $n_s$ and $r_\mathrm{t/s}$ lie within the Planck constraints 
for the following ranges of parameter value: $0.0155 \leq \gamma \leq 0.0196$ and this behaviour is depicted in Fig.~\ref{plot_observable}.

Now we are in a position to solve for the electromagnetic perturbation  by using the solution of $\tilde{v}(k,\eta)$. 
By using the dimensionless variables 
$x = k\eta$, $\bar{A}_{\pm}(k,\eta) = \sqrt{k}A_{\pm}(k,\eta)$ and $V(k,\eta) = \sqrt{k}\tilde{v}(k,\eta)$, and using 
Eqs.(\ref{f1}) and (\ref{coupling functions}), the electromagnetic equation (\ref{eom_A_FM}) becomes,
\begin{equation}
{\cal A}_1 \frac{d^2\bar{A}_{\pm}}{dx^2} + {\cal A}_2 \frac{d\bar{A}_{\pm}}{dx}
\pm {\cal A}_3 \bar{A}_{\pm} = {\cal S}_{\pm} ,
\label{eom_A_FM_dimensionless variable}
\end{equation}
where, the coefficients are,
\begin{eqnarray}
{\cal A}_1 &=& \frac{1}{a_f^2}\bigg(\frac{k}{H}\bigg)^{4+m} 
+ 2(-1)^m x^{4+m} ~~;~~{\cal A}_2 = -\frac{2}{x^2 a_f^2} \bigg(\frac{k}{H}\bigg)^{4+m} 
+ (4 + 2m)(-1)^m x^{3+m} \\
{\cal A}_3 &=& \frac{1}{a_f^2}\bigg(\frac{k}{H}\bigg)^{4+m} 
+ 2(-1)^r(2+r)\bigg(\frac{k}{H}\bigg)^{m-r} x^{3+r} - 8s(-1)^s\bigg(\frac{k}{H}\bigg)^{m-s+2} x^{s+1} \\
{\cal S}_{\pm} &=& \pm 2\sqrt{2}\bigg(\frac{h_0 x^5}{H}\bigg)\bigg(\frac{k}{H}\bigg)^{m-1} \Bigg[(5 + 2\gamma) \bigg(\frac{dV}{dx} + \bigg(\frac{1+\gamma}{x}\bigg)V\bigg)
+ x\bigg(\frac{d^2V}{dx^2} - \frac{(2+3\gamma)}{x^2}V\bigg)\Bigg] .
\end{eqnarray}
Here it may be mentioned that due to the complicated nature, we solve the above equation numerically. For this we will consider 
values of $m$, $r$ and $s$ which lead to a scale invariant magnetic 
power spectrum in the superhorizon limit. For the modes lying outside the horizon, the scalar Mukhanov-Sasaki variable 
behaves as $\tilde{v}(k,\eta) \sim (-k\eta)^{-\nu + \frac{1}{2}} = (-k\eta)^{-1-\gamma}$ (see Eq.(\ref{superhorizon form})), which indicates that 
$\tilde{v}(k,\eta)$ becomes proportional to $z(\eta)$ at the scale $-k\eta \rightarrow 0$. Furthermore the factor $\frac{k}{H}$ present in 
Eq.(\ref{eom_A_FM_dimensionless variable}) is considered to be less than unity, which is indeed true for the CMB scale modes 
$\sim 10^{-40}\mathrm{GeV}^{-1}$ - the scale of interest to estimate the magnetic strength at the current universe. In account of these 
arguments, Eq.(\ref{eom_A_FM_dimensionless variable}) in the superhorizon scale can be written as,
\begin{eqnarray}
 \frac{d^2\bar{A}_{\pm}}{dx^2}&+&\bigg(\frac{2+m}{x}\bigg)\frac{d\bar{A}_{\pm}}{dx}\nonumber\\ 
 &\pm&\left(-1\right)^{r-m}\left\{\left(\frac{k}{H}\right)^{m-r}
 \left(\frac{2+r}{x^{1+m-r}}\right) - 4s\left(-1\right)^{-2-r}\left(\frac{k}{H}\right)^{2+m-s}~x^{3+m-s}\right\}\bar{A}_{\pm}  = 0
 \label{eom_A_superhorizon-new}
\end{eqnarray}
where, it may be observed that ${\cal S}_{\pm}$ vanishes due to the superhorizon solution of $\tilde{v}(k,\eta)$ given in Eq.(\ref{final solution_sp_FM}). 
The above equation may not be solved analytically, thus we consider $m = r \geq s$ (later, we will show that such choices of parameters 
lead to the scale invariant electric as well as the scale invariant magnetic power spectrum). In effect, the term containing 
$\left(k/H\right)^{2+m-s}$ within the curly bracket becomes subdominant compared to the other one, 
since $k/H \ll 1$ for the CMB scale modes, in particular $k/H \sim 10^{-54}$ for $H \sim 10^{14}\mathrm{GeV}$. As a consequence, 
Eq.(\ref{eom_A_superhorizon-new}) turns out to be
\begin{eqnarray}
 \frac{d^2\bar{A}_{\pm}}{dx^2} + \bigg(\frac{2+m}{x}\bigg)\frac{d\bar{A}_{\pm}}{dx} \pm \bigg(\frac{2+m}{x}\bigg)\bar{A}_{\pm} = 0
 \label{eom_A_superhorizon}
\end{eqnarray}
Eq.(\ref{eom_A_superhorizon}) has the following solution for $\bar{A}_{\pm}(k,\eta)$,
\begin{eqnarray}
 \bar{A}_{+}(k,\eta)&=& \big(-k\eta\big)^{\frac{1-m}{2}}\bigg\{C_{1} \mathrm{I}_{-m-1}\big[\sqrt{-(8+4m)x}\big] 
 + D_{1} \mathrm{I}_{m+1}\big[\sqrt{-(8+4m)x}\big]\bigg\}
 \\
  \bar{A}_{-}(k,\eta) &=& \big(-k\eta\big)^{\frac{1-m}{2}}\bigg\{C_{2} \mathrm{I}_{-m-1}\big[\sqrt{(8+4m)x}\big] 
 + D_{2} \mathrm{I}_{m+1}\big[\sqrt{(8+4m)x}\big]\bigg\}
 \label{superhorizon solution A plus}
 \label{superhorizon solution A minus}
\end{eqnarray}
respectively, with $C_{1,2}$ and $D_{1,2}$ are integration constants. Here $\mathrm{I}_n(w)$ symbolizes the modified Bessel function 
of first kind having order $n$ and argument $w$, which, in the limit $w \ll 1$, has a power law form as $\mathrm{I}_n(w) = w^{n/2}$. Thus 
due to the presence of the modified Bessel function in Eq.(\ref{superhorizon solution A plus}), the term containing $C_1$ 
in the expression of $\bar{A}_{+}$ dominate over the other one containing $D_1$, i.e we may write 
$\lim_{|k\eta| \ll 1}\bar{A}_{+} \approx C_1(-k\eta)^{-m}$. Similar argument can also be drawn for $\bar{A}_{-}$, i.e 
$\lim_{|k\eta| \ll 1}\bar{A}_{-} \approx C_2(-k\eta)^{-m}$. As a result, Eq.(\ref{power spectra em}) immediately leads to the electric 
and magnetic spectra in the superhorizon limit, as follows:
\begin{eqnarray}
P^{(E)}(k,\eta)&=&\frac{m^2}{2\pi^2}\big\{|C_1|^2 + |C_2|^2\big\}\bigg[f_1(\eta) + \frac{6f_2(\eta)}{a^2}\bigg]
H^4\big(-k\eta\big)^{2-2m}\nonumber\\
P^{(B)}(k,\eta)&=&\frac{f_1(\eta)}{2\pi^2}\big\{|C_1|^2 + |C_2|^2\big\}H^4\big(-k\eta\big)^{4-2m}~~~.
\label{superhorizon magnetic power spectrum}
\end{eqnarray}
Furthermore, as the two polarization modes evolves differently the magnetic field generated through 
this mechanism will be helical in nature and the helicity power spectrum during inflation comes as,
\begin{eqnarray}
 P^{(h)}(k,\eta) &=& \bigg(\frac{k^3}{2\pi^2a^3}\bigg)\big(-k\eta\big)^{-2m}\bigg\{\big|C_{1}\big|^2 - \big|C_{2}\big|^2\bigg\}~~.
 \label{helicity power spectrum}
\end{eqnarray}
Specifically $(-k\eta)^{-2m}$ factor in the above expression clearly depicts that the comoving helicity spectra increases with 
time for a given mode for any positive value of $m$ during the 
inflationary era.\\ 
Eq.(\ref{superhorizon magnetic power spectrum}) points out that the scale dependence of the electric and the magnetic power spectra are different, 
in particular, $P^{(E)} \sim (-k\eta)^{2-2m}$ and $P^{(B)} \sim (-k\eta)^{4-2m}$. Thereby the electric power spectrum becomes 
scale invariant for $m = 1$ while, in the 
magnetic case, $m = 2$ leads to a scale invariant power spectrum. 
Thereby the coupling exponents consistent 
with a scale invariant magnetic power spectrum are constrained by, $m = 2$, $r = 2$ and $s \leq 2$; while for the scale invariant electric power spectrum, 
the corresponding constraint is given by $m = 1$, $r = 1$ and $s \leq 1$. Hence, in the following, 
we will solve the electromagnetic mode functions and will discuss the possible implications 
for the two different cases.

Here we would like to mention that in the Ratra like model where the EM Lagrangian is 
$\mathcal{L}_{em} = \sqrt{-g}f\left(\phi\right)F_{\mu\nu}F^{\mu\nu}$ with $f(\phi(\eta)) = \left(a(\eta)/a_f\right)^n$, the magnetic and electric power 
spectra become scale invariant for $n = -6$ and $n = -4$ respectively \cite{Haque:2020bip,Kobayashi:2019uqs}. However, 
EFT framework provides us wider possibility of having such power spectrum. 
Assuming the canonical gauge kinetic term to be $f_1(\eta) = \left(a(\eta)/a_f\right)^2$, 
the other EFT coupling functions having exponents $m$, $r$, $s$ (as given in Eq.(\ref{coupling functions})), 
satisfying the condition $s \leq m = r = 2$ can lead to a scale invariant magnetic spectrum, 
while  $s \leq m = r = 1$ will lead to scale invariant electric power spectrum. Therefore, the EFT coupling functions seem to play 
an important role in making the scale invariance of magnetic or electric power spectra in the present context.

Before we embark on estimating the magnetic field strength at large scale in the next section, we would like to address the 
strong coupling problem for general values of $m$, $r$ and $s$ in the 
spirit of EFT. For this purpose we need to start with the action of the EM and Dirac field, which in the present context, can be expressed by,
\begin{eqnarray}
 S_{em} + S_{\Psi}&=&\int d^4x\sqrt{-g}\bigg[-\frac{1}{4}f_1(\eta) F_{\mu\nu}F^{\mu\nu} + f_2(\eta)F^0_{i}F^{0i} + f_3(\eta)\epsilon^{ijk}F^0_iF_{jk} 
 + f_4(\eta)\epsilon^{\mu\nu\alpha\beta}F_{\mu\nu}F_{\alpha\beta}\bigg]\nonumber\\
 &+&\int d^4 x \sqrt{-g}\bigg[\alpha(\eta)\bar{\Psi}~i~\gamma^{\mu}\mathbf{D}_{\mu}\Psi~+~ \beta(\eta)\bar{\Psi}~i~\gamma^0\mathbf{D}_0\Psi\bigg]
 \label{strong coupling full action}
\end{eqnarray}
where $\Psi$ is the Dirac field and $\mathbf{D}_{\mu} = \nabla_{\mu} - ieA_{\mu}$ is the covariant derivative. The part of the EM field in the above action 
has been considered earlier in Eq.(\ref{em action}), however in regard to the Dirac field, $S_{\Psi}$ contains two terms with EFT coefficients 
$\alpha(\eta)$ and $\beta(\eta)$. The underlying spatial diffeomorphism symmetry allows the coefficients $\alpha(\eta)$ and $\beta(\eta)$ 
to depend on $\eta$ only. Moreover the term $\beta(\eta)\bar{\Psi}i\gamma^0\mathbf{D}_0\Psi$ behaves as a scalar quantity under 
spatial diffeomorphism symmetry transformation and thus can be a possible term in $S_{\Psi}$ in the context of EFT of cosmology \cite{Tasinato:2014fia}. 
By expanding the $S_{\Psi}$ in terms of temporal and spatial parts over the FRW spacetime, we get,
\begin{eqnarray}
S_{\Psi}=\int d^4 x a^4 \bigg[i(\alpha(\eta)+\beta(\eta))\bar{\Psi}\gamma^0 \mathbf{D}_0 \Psi +i\alpha(\eta)\bar{\Psi}\gamma^i \mathbf{D}_i \Psi\bigg]
\label{sc dirac kinetic}
\end{eqnarray}
which makes Dirac field  non-canonical. Therefore in order to make the Dirac field 
canonical, we redefine the field as,
\begin{eqnarray}
\Psi \longrightarrow \widetilde{\Psi} = a^2 \sqrt{\alpha(\eta)+\beta(\eta)} ~\Psi~~.
\label{normalized Dirac}
\end{eqnarray}
Similarly due to the presence of the coupling functions $f_i(\eta)$ ($i = 1,2,3,4$), the electromagnetic kinetic part becomes non-canonical, 
in particular,
\begin{eqnarray}
 S_{em}^{(kin)} = \frac{1}{2}\int d^4 x \bigg[\left(f_1(\eta)+\frac{2 f_2(\eta)}{a^2}\right)(A_i^{\prime})^2\bigg]~~.
\end{eqnarray}
Thus we redefine the EM field as,
\begin{eqnarray}
 A_i \longrightarrow \widetilde{A}_i = \left(f_1(\eta)+ \frac{2 f_2(\eta)}{a^2}\right) A_i
 \label{normailzed EM}
\end{eqnarray}
which in turn restores the canonicity of the EM field. In terms of such canonical variables, Eq.(\ref{strong coupling full action}) immediately leads 
to the U(1) interaction between the EM and the Dirac fields as, 
\begin{eqnarray}\label{normalized_int}
S_{int} = \int d^4 x \bigg[\frac{e~ \alpha(\eta)}{(\alpha(\eta)+\beta(\eta))\sqrt{f_1(\eta)+\frac{2 f_2(\eta)}{a^2}}}\bigg]
\bar{\widetilde{\Psi}} i~\gamma^i \widetilde{A}_i \widetilde{\Psi} 
= \int d^4 x ~\big(e~\lambda_\mathrm{eff}\big)~ \bar{\widetilde{\Psi}} i~\gamma^i \widetilde{A}_i \widetilde{\Psi}
\end{eqnarray}
here $\lambda_\mathrm{eff}$ defines an effective electric charge as
\begin{eqnarray}\label{lambda}
e_{eff} = e \lambda_\mathrm{eff} = \frac{e}{{\left(1+\frac{\beta(\eta)}{\alpha(\eta)}\right)\sqrt{f_1(\eta)+\frac{2 f_2(\eta)}{a^2}}}}
\end{eqnarray}
In order to avoid the strong coupling problem we need to have $\lambda_\mathrm{eff} \leqslant 1$ which leads to the following condition on  the 
EFT parameters,  
\begin{eqnarray}\label{sc_constraint}
\left(1+\frac{\beta(\eta)}{\alpha(\eta)}\right)\geqslant \frac{1}{\sqrt{f_1(\eta)+\frac{2 f_2(\eta)}{a^2}}}~~.
\end{eqnarray}
Thus, in the context of EFT magnetogenesis, the strong coupling problem can be naturally resolved provided the functions 
$a(\eta)$, $\beta(\eta)$, $f_1(\eta)$ and $f_2(\eta)$ satisfy Eq.(\ref{sc_constraint}) during inflation. In the present work, we consider 
$f_1(\eta) = \big(a/a_\mathrm{f}\big)^2$ and $f_2(\eta) = \big(a_\mathrm{i}/a\big)^m$ (see Eqs.(\ref{f1}) and (\ref{coupling functions})), 
due to which the above equation becomes,
\begin{eqnarray}\label{final_sc_constraints}
\left(1+\frac{\beta(\eta)}{\alpha(\eta)}\right)~\geqslant~\frac{1}{\sqrt{\big(\frac{a}{a_f}\big)^2+ \frac{2 f_2}{a^2}\big(\frac{a_i}{a}\big)^m}}~~.
\end{eqnarray}
As mentioned earlier, we are interested in two distinct cases depending on whether the magnetic power spectrum or electric power spectrum becomes 
scale invariant, for which the corresponding parametric regime are given by 
$m = 2$, $r = 2$, $s \leq 2$ or $m = 1$, $r = 1$, $s \leq 1$. Thereby depending on such values of $m$ along with 
$a(\eta) = (-H\eta)^{-1}$, the functions $\alpha(\eta)$ and $\beta(\eta)$ can be chosen properly so that Eq.(\ref{final_sc_constraints}) holds true 
during inflation, which in turn leads to the resolution of the strong coupling problem.

\section{Case-I: Scale invariant magnetic power spectrum}\label{sec_case-I}

\begin{figure}[t]
\begin{center}
 \centering
 \includegraphics[width=3.5in,height=2.5in]{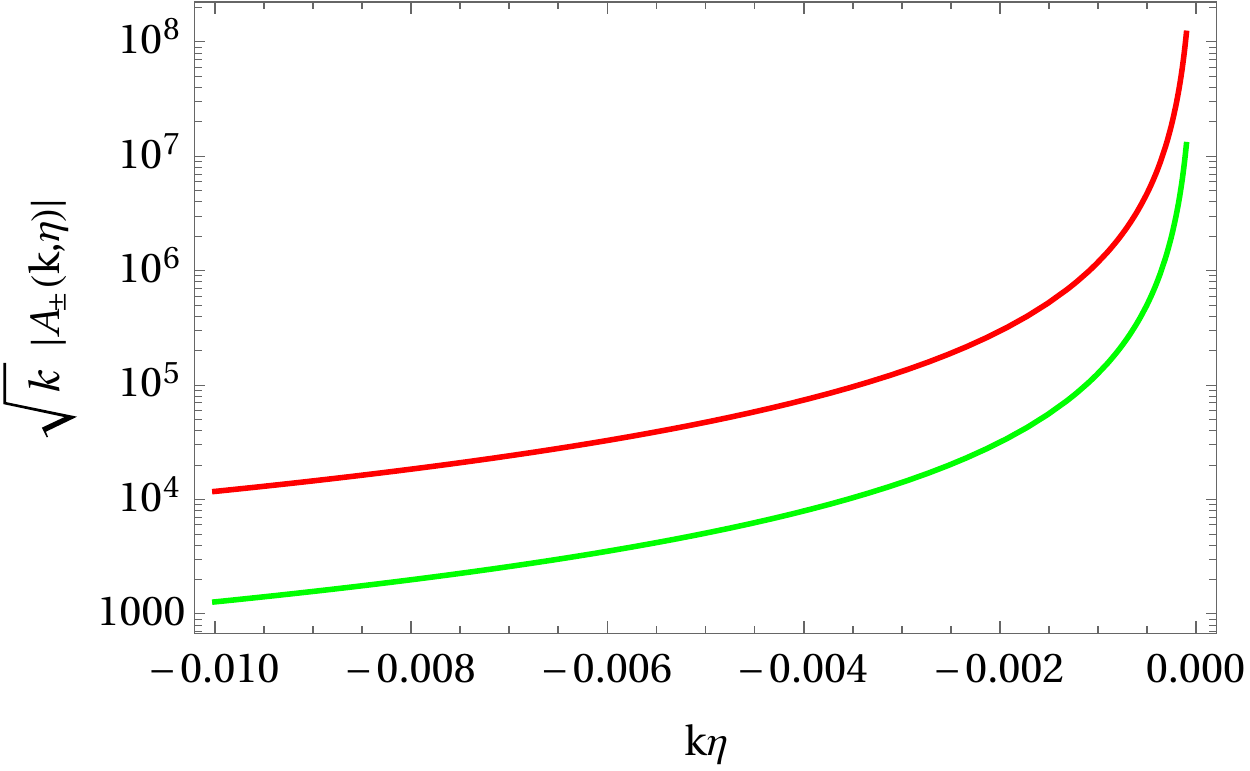}
 \caption{The red line represents $\sqrt{k}A_{+}(k,\eta)$ versus $k\eta$ and the green line represents $\sqrt{k}A_{-}(k,\eta)$ 
 versus $k\eta$. Both the plots are for $k = 0.05\mathrm{Mpc}^{-1} \approx 2\times10^{-40}\mathrm{GeV}$, $H = 10^{14}\mathrm{GeV}$ and 
 $m=r=s=2$ respectively.}
 \label{plot_em_numerical_solution}
\end{center}
\end{figure}

In this section, we consider the case where the magnetic power spectrum is scale invariant for the choices of the parameters $m = 2$, $r = 2$ 
and $s \leq 2$. Considering these set of parameter values we numerically 
solve Eq.(\ref{eom_A_FM_dimensionless variable}) for $\bar{A}_{\pm}(k,\eta)$ which are depicted in Fig.\ref{plot_em_numerical_solution}. 
During the numerical analysis, 
we take the Bunch-Davies initial condition for the electromagnetic mode functions 
and the initial conditions are set for the modes when they are well within the horizon length scale (i.e $-k\eta = 10$). 
Moreover the numerical solutions are obtained upto the epoch when the 
modes are well outside the horizon ($-k\eta = 10^{-4}$ has been used in our calculation).

Fig.[\ref{plot_em_numerical_solution}] clearly demonstrates that the $A_{+}$ mode dominates over 
the $A_{-}$ mode and thus the electromagnetic energy density in superhorizon limit is effectively determined by that of 
the contribution of $A_{+}$ mode.

\subsection{Present magnetic strength and backreaction problem}\label{sec_present_magnetic_strength1}
In this section we will concentrate on the strength of the magnetic field in the present epoch. 
Conventionally one considers
high electrical conductivity in the post inflationary epoch  within the instantaneous reheating scenario, where inflaton makes a sudden 
jump from the inflationary phase to a radiation dominated phase.  
Due to large electrical conductivity electric field quickly vanishes and magnetic field freezes which will then 
adiabatically evolve till today in the form of large scale magnetic field. As we mentioned earlier for our analysis,  
coupling functions $f_1(\eta) = 1$, $f_2(\eta) \approx f_3(\eta) \approx f_4(\eta) \approx h(\eta) \approx 0$, are assumed which will restore 
conformal symmetry and  electromagnetic field follows the standard Maxwell's equations. 
Hence, the frozen in magnetic field energy density evolves as $1/a^4$. The magnetic field strength at the present epoch is related 
with that at the end of inflation by the expression
 \begin{eqnarray}
 \frac{\partial \rho(\vec{B})}{\partial \ln{k}}\bigg|_{0} = \bigg(\frac{a_f}{a_0}\bigg)^4~\frac{\partial \rho(\vec{B})}{\partial \ln{k}}\bigg|_{{\eta_f}},
 \label{magnetic strength 1}
\end{eqnarray}
where $\eta_f$ is the conformal time at the end of inflation and the suffix '0' denotes present time. With Eq.(\ref{power spectra em}), 
the above equation yields the present magnetic strength ($B_0$) 
\begin{eqnarray}
 B_0 = \frac{1}{2\pi}~\bigg\{\sqrt{\big|\bar{A}_{+}(k,\eta_f)\big|^2 + \big|\bar{A}_{-}(k,\eta_f)\big|^2}\bigg\}
 \bigg(\frac{a_f}{a_0}\bigg)^2~H^2\big(-k\eta_f\big)^{2},
 \label{magnetic strength 2}
\end{eqnarray}
where we recall that $\bar{A}_{\pm} = \sqrt{k}A_{\pm}$ and $k$ denotes the CMB scale ($\simeq 0.02\mathrm{Mpc}^{-1}$) 
at which we will estimate the current magnetic strength. In order to estimate $B_0$ from Eq.(\ref{magnetic strength 2}), 
we need to know $\frac{a_0}{a_f}$. The factor $\frac{a_0}{a_f}$ can be determined from the entropy conservation relation, i.e. from 
$gT^3a^3 = \mathrm{constant}$, where $g$ is the effective relativistic degrees of freedom and $T$ denotes the temperature of the relativistic fluid, 
which finally yields $\frac{a_0}{a_f} \approx 10^{30}\big(H/10^{-5}M_{Pl}\big)^{1/2}$, with $H$ being the 
Hubble parameter during inflation and, in particular, we consider $H = 10^{-5}M_{Pl}$. Moreover by using the numerical solution 
of $\bar{A}_{\pm}k,\eta)$ from Fig.[\ref{plot_em_numerical_solution}], we get, $\bar{A}_{+}(k,\eta_f) \simeq 10^{8}$ and 
$\bar{A}_{-}(k,\eta_f) \simeq 10^{7}$ for $-k\eta_f = 10^{-4}$. With such ingredients of $\frac{a_0}{a_f}$ and $\bar{A}_{\pm}k,\eta_f)$, 
we estimate the magnetic strength at the present epoch from Eq.(\ref{magnetic strength 2}) to be
\begin{eqnarray}
 B_0\bigg|_{CMB} = 0.82\times10^{-13}\mathrm{G},
 \label{magnetic strength 3}
\end{eqnarray}
where we use the conversion $1\mathrm{G} = 1.95\times10^{-20}\mathrm{GeV}^2$. 
The above result gives us a typical value for the magnetic field at the present epoch in our present framework. It may be noticed that 
$B_0$ indeed depends on the inflationary Hubble parameter, in particular if we consider $H = 10^{-6}M_{Pl}$, then $B_0$ will be estimated as 
$B_0 \simeq 10^{-14}\mathrm{G}$ which, in comparison with Eq.(\ref{magnetic strength 3}), clearly argue that the 
current magnetic strength for $H = 10^{-6}M_{Pl}$ gets 1 order lower than that of the case where $H = 10^{-5}M_{Pl}$. 
However, from the observational \cite{Widrow:2002ud,Kandus:2010nw,Durrer:2013pga} results a constraint on the current magnetic strength of 
$10^{-10}G \lesssim B_0 \lesssim 10^{-22}G$ is obtained around the CMB scales. Therefore the theoretical 
prediction of $B_0$ lies within the observational constraints in the case where the magnetic power spectrum turns out to be scale invariant.\\
At this stage it deserves mentioning that the present case (i.e $m = 2$), despite having compatibility with the observational strength of 
$B_0$, hinges with the 
backreaction problem. In particular, for $m = 2$, we have from Eq.(\ref{superhorizon magnetic power spectrum}) $P^{(E)}(k,\eta) \propto (-k\eta)^{-2}$. 
In this case, as $-k\eta \rightarrow 0$ (i.e. towards the end of inflation), the electric energy density increases rapidly 
as $P^{(E)} \propto (-k\eta)^{-2} \rightarrow \infty$. Thereby the model runs into {\em difficulties} as the electric energy density would eventually 
exceed the inflaton energy density in the universe even before inflation ends. 

Thus the scenario, where the magnetic power spectrum becomes scale invariant,  can lead to required value of magnetic strength, however 
suffers from the backreaction problem. On contrary, the situation becomes different in the case where the electric power 
spectrum becomes scale invariant (i.e for $m = 1$), as discussed in the following section.\\

\section{Case-II: Scale invariant electric power spectrum}\label{sec_case-II}

\begin{figure}[t]
\begin{center}
 \centering
 \includegraphics[width=3.5in,height=2.5in]{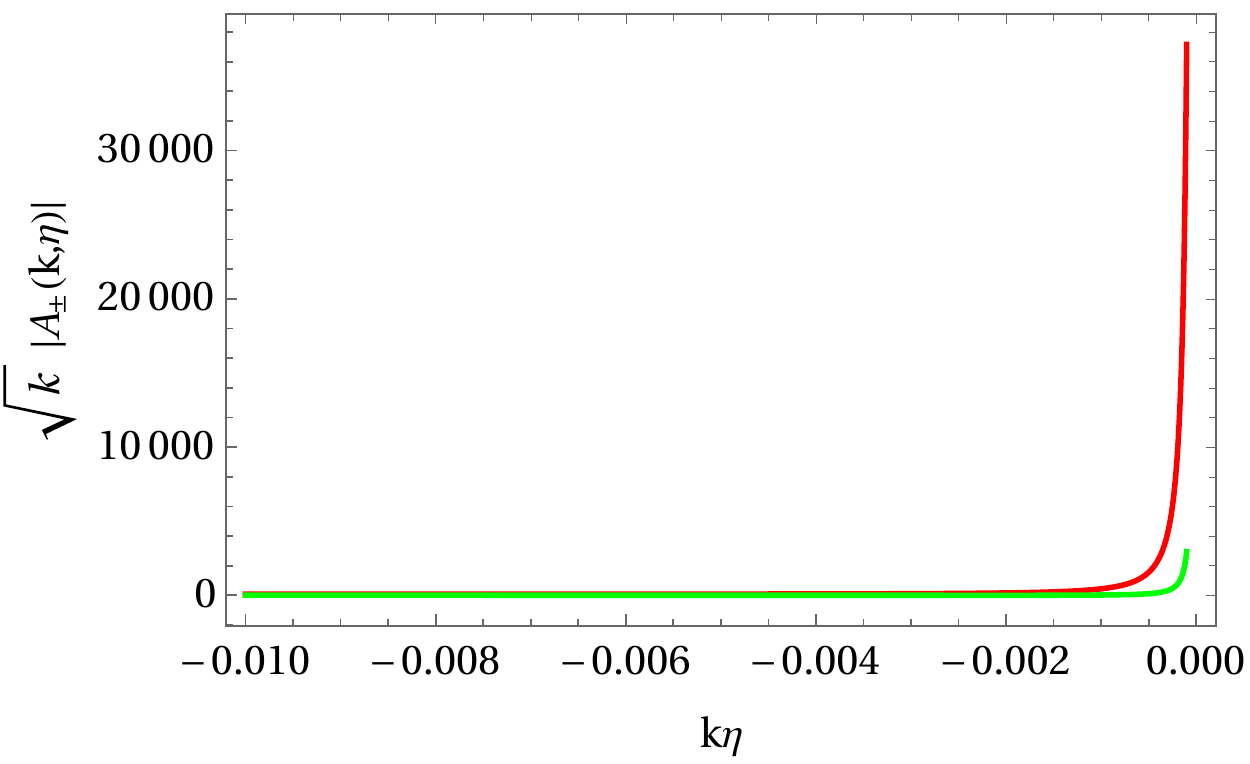}
 \caption{The red line represents $\sqrt{k}A_{+}(k,\eta)$ versus $k\eta$ and the green line represents $\sqrt{k}A_{-}(k,\eta)$ 
 versus $k\eta$. Both the plots are for $k = 0.05\mathrm{Mpc}^{-1} \approx 2\times10^{-40}\mathrm{GeV}$, $H = 10^{14}\mathrm{GeV}$ and 
 $m=r=s=1$ respectively.}
 \label{plot_em_numerical_solution-II}
\end{center}
\end{figure}

\begin{figure}[ht]
\begin{center}
 \centering
 \includegraphics[width=3.5in,height=2.5in]{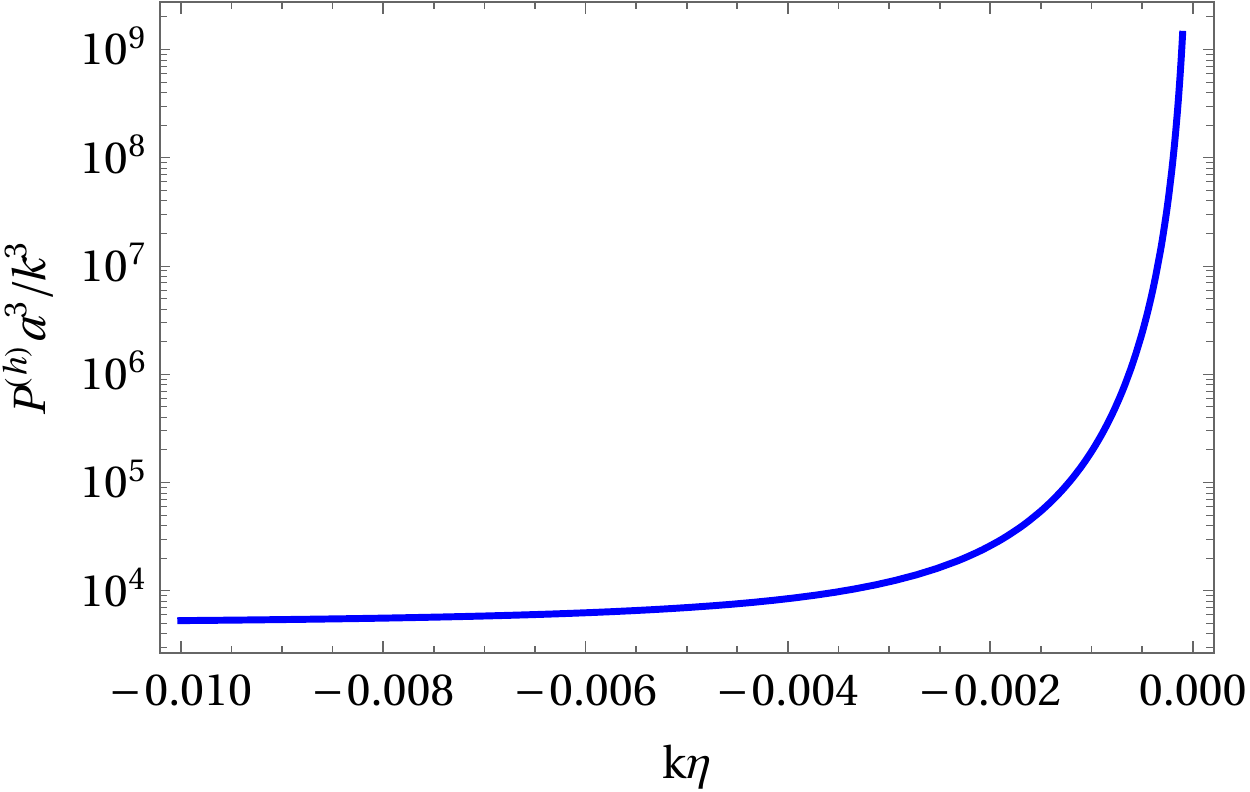}
 \caption{$P^{(h)}a^3/k^3$ versus $k\eta$ during inflationary epoch.}
 \label{plot_helicity_inflation}
\end{center}
\end{figure}

From Eq.(\ref{superhorizon magnetic power spectrum}) it is evident that the electric power spectrum becomes 
scale invariant for the parametric space satisfying 
$s \leq m = r = 1$. Thereby considering a particular set of parameter values, in particular $m=r=s=1$, we solve the electromagnetic mode functions 
from Eq.(\ref{eom_A_FM_dimensionless variable}) with the Bunch-Davies initial condition. 
However due to the complicated nature, Eq.(\ref{eom_A_FM_dimensionless variable}) is solved 
numerically and they are depicted in Fig.[\ref{plot_em_numerical_solution-II}]. The numerical analysis is 
performed for $k = 0.05\mathrm{Mpc}^{-1}$ within the same same range $ -10 < k\eta < 10^{-4}$ from sub to super Hubble scale.\\
Fig.\ref{plot_em_numerical_solution-II} depicts that similar to the previous case, the $A_{+}$ mode is slightly dominated over the 
$A_{-}$ mode towards the end of inflation 
and thus the electromagnetic energy density in superhorizon limit is effectively determined by that of the contribution of $A_{+}$ mode. 
The solutions of $A_{\pm}$ immediately lead to the evolution of the helicity density during inflationary era. This is 
depicted in Fig.\ref{plot_helicity_inflation} which clearly demonstrates that the comoving helicity density during inflation increases with time.

The scenario, where the electric power spectrum becomes scale invariant (i.e for $s\leq m=r=1$), leads to 
the electric and magnetic power spectra in the superhorizon scale as,
\begin{eqnarray}
P^{(E)}(k,\eta)&=&\frac{m^2}{2\pi^2}H^4\big\{|C_1|^2 + |C_2|^2\big\}\bigg[f_1(\eta) + \frac{6f_2(\eta)}{a^2}\bigg]\nonumber\\
P^{(B)}(k,\eta)&=&\frac{f_1(\eta)}{2\pi^2}H^4\big\{|C_1|^2 + |C_2|^2\big\}\big(-k\eta\big)^{2}~~~.
 \label{superhorizon magnetic power spectrum case-II}
\end{eqnarray}
respectively, where we use Eq.(\ref{superhorizon magnetic power spectrum}). 
Thereby the spectral indices of the electric and magnetic spectra, defined by $n(E) = \frac{\partial\ln{P^{(E)}}}{\partial\ln{k}}$ 
and $n(B) = \frac{\partial\ln{P^{(B)}}}{\partial\ln{k}}$, become $n(E) = 0$ and $n(B) = 2$ respectively. It may be observed that 
in the scenario where the electric power spectrum becomes scale invariant, the spectral index for the magnetic power becomes positive, which in turn 
hints to the resolution of the backreaction problem. 
However, in order to investigate whether the backreaction of the electromagnetic field stays small during inflation, 
we shall first consider the energy stored in the electric field at a given time $\eta_c$, which is 
\begin{eqnarray}
 \rho(\vec{E},\eta_c) = \int_{k_i}^{k_c} \frac{\partial \rho(\vec{E})}{\partial \ln{k}} d\ln{k} 
 = \frac{1}{2\pi^2}\bigg[e^{-2\big(N_f - N_c\big)} + 6e^{-3N_c}\bigg]\big\{|C_1|^2 + |C_2|^2\big\}H^4N_c~~,
 \label{electric energy density 1}
\end{eqnarray}
here $k_i$ and $k_c$ denote the modes that cross the horizon at the beginning of inflation and at $\eta = \eta_c$, respectively, and thus $k_i$ 
is considered to be the CMB scale. Moreover, $N_f$ is the total e-folding number of the 
inflation era. $N_c = \ln{\big(\frac{a_c}{a_i}\big)}$, with 
$a_i = a(\eta_i)$ and $a_c = a(\eta_c)$, 
is the e-fold number associated with conformal time $\eta = \eta_c$  during inflation. 
We also have the relation $|\eta_c| = k_c^{-1}$ and $k_c$ is obviously greater than the CMB mode 
momentum $k_{CMB} \approx 10^{-40}\mathrm{GeV} \approx 0.05 \mathrm{Mpc}^{-1}$. Similarly, we determine 
the energy density coming from the magnetic fields, which yields
\begin{eqnarray}
 \rho(\vec{B},\eta_c) = \int_{k_i}^{k_c} \frac{\partial \rho(\vec{B})}{\partial \ln{k}} d\ln{k} 
 = \frac{1}{4\pi^2}e^{-2\big(N_f - N_c\big)}\big\{|C_1|^2 + |C_2|^2\big\}H^4\bigg(1 - e^{-2N_c}\bigg)~~.
 \label{magnetic energy density 1}
\end{eqnarray}
To derive the above two expressions, we use the form of the coupling functions at $\eta = \eta_c$ in terms of the e-folding number, as 
$f_1(\eta_c) = \big(a_c/a_f\big)^2 = e^{-2\big(N_f - N_c\big)}$ and $f_2(\eta) = \big(a_i/a_c\big) = e^{-N_c}$. 
The total electromagnetic energy density at $\eta = \eta_c$ becomes
\begin{eqnarray}
 \rho_{em}(\eta_c) = \rho(\vec{E},\eta_c) + \rho(\vec{B},\eta_c) 
 &=&\frac{1}{4\pi^2}\big\{|C_1|^2 + |C_2|^2\big\}H^4\bigg\{N_c\left(2e^{-2\big(N_f - N_c\big)} + 12e^{-3N_c}\right)\nonumber\\ 
 &+&e^{-2\big(N_f - N_c\big)}\left(1 - e^{-2N_c}\right)\bigg\}
 \label{total energy density}
\end{eqnarray}
where $C_{1,2}$ are integration constants as mentioned in Eq.(\ref{superhorizon magnetic power spectrum}). 
In order to avoid the backreaction issue, we have to ensure that the 
electromagnetic energy density is less than that of the background energy density. In particular, we have to show $\rho_{em} < 3M_{Pl}^2H^2$ during  
inflation. The above expression, along with the consideration of $N_c \leq N_f \approx 60$, clearly indicate that the 
electromagnetic energy density during inflation is of the order of $H^4$, i.e.
\begin{eqnarray}
 \rho_{em}(\eta_c) \sim H^4~~~.
 \label{backreaction}
\end{eqnarray}
The inflationary energy scale is less than the Planck scale; in particular, we consider $H = 10^{-5}M_{Pl}$ and, thus, Eq.(\ref{backreaction}) leads to the 
inequality  $\rho_{em} \ll M_{Pl}^2H^2$. This confirms that the electromagnetic field has a negligible backreaction 
on the background inflationary spacetime, leading to the resolution of the backreaction problem in the present magnetogenesis scenario.\\

\subsection{Present magnetic strength}

Following same methodology discussed for the scale invariant magnetic case (Sec.[\ref{sec_present_magnetic_strength1}]), 
and using Eqs.(\ref{superhorizon solution A plus}) with $m = 1$, 
the expression of the current magnetic strength turns out to be, 
\begin{eqnarray}
 B_0 = \frac{1}{2\pi}~\bigg\{\sqrt{\big|C_1\big|^2 + \big|C_2\big|^2}\bigg\}\bigg(\frac{a_f}{a_0}\bigg)^2~H^2\big(-k\eta_f\big),
 \label{magnetic strength 3-II}
\end{eqnarray}
Here $C_{1,2}$ are integration constants appeared in Eq.(\ref{superhorizon magnetic power spectrum}) and 
$k$ symbolizes the momentum of the CMB scale $\approx 0.05\mathrm{Mpc}^{-1}$. As discussed earlier, by using the entropy conservation we get 
$\frac{a_0}{a_f} = 10^{30}\big(H/10^{-5}M_{Pl}\big)^{1/2}$ with $H$ being the inflationary Hubble scale. Moreover 
in regard to $-k\eta_f$, we have the relation $|\eta_f| = \frac{1}{k_f}$ where $k_f$ is the mode which crosses the horizon at the end of inflation and 
thus $-k\eta_f$ is given by $-k\eta_f = \exp{\big(N_f\big)}$. With such expressions of $\frac{a_0}{a_f}$ and $-k\eta_f$, along with $H = 10^{-5}M_{Pl}$, 
we estimate the magnetic strength at the present epoch from Eq.(\ref{magnetic strength 3-II}) to be,
\begin{eqnarray}
 B_0\bigg|_{CMB} \sim 10^{-40}\mathrm{G}
 \label{magnetic strength 4-II}
\end{eqnarray}
where we use $N_f = 60$ and the conversion $1\mathrm{G} = 1.95\times10^{-20}\mathrm{GeV}^2$ may be useful. 
The above result provides an estimation for the magnetic field 
at the present epoch, as obtained from the framework. However, the observational results put a constraint on the 
current magnetic strength as $10^{-10}\mathrm{G} \lesssim B_0 \lesssim 10^{-22}\mathrm{G}$ around the CMB scales. Therefore it is clear from 
Eq.(\ref{magnetic strength 4-II}) that the theoretical prediction of $B_0$ lies far below than the range of the observational expectation.

Thus as a whole, by comparing Secs.[\ref{sec_case-I}] and [\ref{sec_case-II}], it is evident that 
the scenario where the magnetic power spectrum is scale invariant (let, call it S1) gets distinct features in comparison to 
that of where the electric spectrum is scale invariant (let, call it S2). In particular, the scenario S1 is found to be 
observationally compatible with regard to the present magnetic strength, however suffers from the backreaction problem. On the other hand, 
the scenario S2 is indeed free from 
the backreaction problem, but does not predict sufficient magnetic field strength in present universe. Such distinctions of S1 and S2 are 
clearly depicted in Table[\ref{Table-1}]. However both the scenarios S1 and S2 are able to resolve the strong coupling problem as indicated earlier 
in Eq.(\ref{final_sc_constraints}).
\\
 \begin{table}[t]
  \centering
  \begin{tabular}{|c|c|c|}
   \hline 
   Scenario & First feature  & Second feature\\
   \hline
   S1 ($P^{(B)}$ is scale invariant) & Compatible with observation& Suffers from backreaction problem \\
   \hline
   S2 ($P^{(E)}$ is scale invariant) & Not compatible with observation & Free from backreaction problem \\
   \hline
   \hline
  \end{tabular}%
  \caption{Comparison between the scenarios S1 and S2}
  \label{Table-1}
 \end{table}
Above issues motivated us to look for physical solutions which can be theoretically as well as observationally viable concurrently. 
And for this we would consider the case when reheating play an important role. It deserves mentioning that our discussions on scenarios 
S1 and S2 are based on the fact that radiation domination starts immediately after the inflation ends which we call  $instantaneous~reheating$ case. 
However a more physical approach would be if, during the cosmic evolution, the universe experiences a reheating phase with $non~zero$ e-fold number. 
In the following sections, we will discuss this possibility and its implications. 
We further assume that during reheating the conductivity will be vanishingly small as it is known that during entire period of reheating dynamics 
is mostly oscillating inflaton dominated.  During the cosmological expansion of the universe, 
the EM field eventually evolves through the reheating phase and thus the electric 
and magnetic field(s) evolutions are considerably affected as compared to the instantaneous reheating case. 
As a consequence, there is a possibility that such an elongated reheating phase may make the models viable 
which we will investigate in Sec.[\ref{sec_reheating}]. 
However, our analysis shows that the presence of a reheating phase will not cure the difficulty of the scenario S1, because S1 suffers 
from the backreaction problem during inflation, which can not be rescued by a reheating phase that occurs after the inflation. Thus, in the 
following we will consider the scenario S2 with an elongated reheating phase and depending on the background reheating dynamics, two cases will arise -- 
(i) when the reheating phase is characterized by the Kamionkowski like mechanism and (ii) where the reheating dynamics follow the perturbative 
mechanism. 
\section{Scenario with scale invariant electric field and reheating phase}\label{sec_reheating}
The presence of a reheating phase with non-zero e-folding number considerably affects the magnetic field evolution in comparison to the instantaneous 
reheating case which we have considered in Sec.[\ref{sec_case-II}]. Actually in the instantaneous reheating scenario, the conductivity becomes huge after the 
end of inflation and as a consequence, the magnetic field evolves as $1/a^4$ from the end of inflation to the present epoch. However, if the reheating 
phase is considered to have a non-zero e-fold number (a natural consideration), then there is no reason to consider a large conductivity immediately 
after the inflation. In particular, the conductivity remains non-zero and small during the reheating epoch and 
consequently, the strong electric field induces the magnetic field \cite{Kobayashi:2019uqs}. 
Due to such Faraday's induction from electric to magnetic field, the redshift of the magnetic field becomes lesser as compared to $1/a^4$ 
in the reheating epoch, and thus the 
magnetic field's present strength may become larger than what has been estimated in Eq.(\ref{magnetic strength 4-II}). 
Motivated by this situation, we aim to determine the magnetic field's current strength 
in the present magnetogenesis model by considering the reheating phase to have a non-zero e-fold number.

To this end let us point out that extended reheating phase is known to  effect the evolution of magnetic field at the 
background level \cite{Demozzi:2009fu,Demozzi:2012wh}. However importance of the Faraday effect during this phase has 
been critically explored recently in the context of magnetogenesis scenario \cite{Kobayashi:2019uqs,Haque:2020bip,Bamba:2020qdj}. 
Furthermore, the effect of reheating is independent of our EFT construction which is only applied during inflation. Most important point 
is that accounting EM Faraday effect can put interesting constraints on the reheating parameters  such as reheating temperature, inflaton 
equation of state, reheating e-folding number. Those constraints in turn will propagate into constraints on the EFT inflationary parameters in a 
model independent way. The constraints on the reheating and EFT inflationary parameters can be observed to be intertwined due to this 
reheating phase. We will discuss these in detail in the following sections.

\subsection{Electromagnetic field evolution and the corresponding power spectrum during reheating}
The effective theory parameters are so chosen, $f_1(\eta) = 1$, $f_2(\eta) = f_3(\eta) = f_4(\eta) = h(\eta) = 0$ 
that after the end of inflation the  conformal symmetry  coupling of the 
electromagnetic field (EM) is restored and, hence, the EM field evolution will follow the standard Maxwell's equations in vacuum \cite{Kobayashi:2019uqs}. 
In particular, the equations of motion of $A_{\pm}(k,\eta)$ are given by,
\begin{eqnarray}
 A_{\pm}''^{(re)}(k,\eta) + k^2A_{\pm}^{(re)}(k,\eta) = 0,
 \label{reheating eom}
\end{eqnarray}
where $A_{\pm}^{(re)}(k,\eta)$ represents the electromagnetic mode function during reheating (i.e the superscript 're' denotes the reheating era). 
As a consequence of the restored conformal symmetry, the further quantum production of the gauge field stops during the reheating stage, 
in particular the absolute value of the Bogoliubov coefficient in the post-inflation becomes time independent. At this stage it deserves mentioning that 
we consider the universe to be a poor conductor, in particular the universe is considered to have a zero conductivity during the reheating era. 
However due to the Schwinger production, such consideration of zero conductivity needs an investigation, which, in the context of EFT, is expected 
to study in near future. Coming back to Eq.(\ref{reheating eom}), the corresponding solution of the mode functions are,
\begin{eqnarray}
 A_{\pm}^{(re)}(k,\eta) = \frac{1}{\sqrt{2k}}\bigg[\alpha_{\pm}(k)~e^{-ik(\eta - \eta_f)} + \beta_{\pm}(k)~e^{ik(\eta - \eta_f)}\bigg],
 \label{reheating solution1}
\end{eqnarray}
with $\alpha_{\pm}(k)$, $\beta_{\pm}(k)$  being two integration constants and $\eta_f$ is the end instant of inflation. The integration constants 
can be determined by matching the EM mode functions and their derivative (with respect to the conformal time) at the junction of inflation-to-reheating; 
in particular,
\begin{eqnarray}
 A_{\pm}^{(re)}(k,\eta_f) = A_{\pm}(k,\eta_f)~~~~~~~~~~~\mathrm{and}~~~~~~~~~~~A_{\pm}'^{(re)}(k,\eta_f) = A_{\pm}'(k,\eta_f),
 \label{reheating continuity1}
\end{eqnarray}
here $A_{\pm}(k,\eta)$ represents the mode function during inflation and follows Eq.(\ref{eom_A_FM}) (note, the EM mode function during inflation 
is symbolized without any superscript, while the EM mode function during reheating is designated by the superscript 're'). Eqs.(\ref{reheating solution1}) 
and (\ref{reheating continuity1}) immediately lead to $\alpha_{\pm}(k)$ and $\beta_{\pm}(k)$ as,
\begin{eqnarray}
 \alpha_{\pm}(k)&=&\sqrt{\frac{k}{2}}~A_{\pm}(k,\eta_f) + \frac{i}{\sqrt{2k}}~A_{\pm}'(k,\eta_f)\nonumber\\
 \beta_{\pm}(k)&=&\sqrt{\frac{k}{2}}~A_{\pm}(k,\eta_f) - \frac{i}{\sqrt{2k}}~A_{\pm}'(k,\eta_f)
 \label{continuity2}
\end{eqnarray}
with $A_{\pm}(k,\eta_f)$ can be obtained from their superhorizon solution as obtained in Eqs.(\ref{superhorizon solution A plus}) and 
(\ref{superhorizon solution A minus}), by putting $\eta = \eta_f$.
If we think in terms of particle production during inflation, the above integration constants $\alpha_{\pm}(k)$ and $\beta_{\pm}(k)$ 
can be identified with the Bogoliubov coefficients at time $\eta = \eta_f$ for incoming and out going mode respectively. 
Naturally those remain constants during the subsequent evolution.
$\big|\beta_k(\eta)\big|$ represents the total number of produced particles (having momentum $k$) 
at time $\eta_f$ from the Bunch-Davies vacuum defined at $\eta \rightarrow -\infty$. Hence, the time independency of the 
Bogoliubov coefficients during the reheating phase is a direct consequence of the fact that the conformal symmetry of the electromagnetic field 
is restored after inflation. 
Nonetheless, by plugging back the solution of $A_{\pm}^{(re)}(k,\eta)$ into Eq.(\ref{power spectra em}) and 
by using the conformal invariance of the electromagnetic field (i.e $f_1(\eta) = 1$, $f_2(\eta) = 0$), 
we determine the magnetic and electric power spectra during the reheating epoch, as follows
\begin{eqnarray}
 \frac{\partial \rho(\vec{B})}{\partial \ln{k}}&=&\frac{1}{2\pi^2}~\sum_{r=\pm}~\frac{k^5}{a^4}\big|A_r^{(re)}(k,\eta)\big|^2\nonumber\\
 &=&\frac{1}{2\pi^2}~\sum_{r=\pm}\bigg(\frac{k^4}{a^4}\bigg)\bigg[|\alpha_r|^2 + |\beta_r|^2 
 + 2 |\alpha_r|~|\beta_r|~\cos{\big\{\theta^{(r)}_1 - \theta^{(r)}_2 - 2k(\eta - \eta_f)\big\}}\bigg]\\
  \frac{\partial \rho(\vec{E})}{\partial \ln{k}}&=&\frac{1}{2\pi^2}~\sum_{r=\pm}~\frac{k^2}{a^4}\big|A_r'^{(re)}(k,\eta)\big|^2\nonumber\\
 &=&\frac{1}{2\pi^2}~\sum_{r=\pm}\bigg(\frac{k^4}{a^4}\bigg)\bigg[|\alpha_r|^2 + |\beta_r|^2 
 - 2|\alpha_r|~|\beta_r|~\cos{\big\{\theta^{(r)}_1 - \theta^{(r)}_2 - 2k(\eta - \eta_f)\big\}}\bigg],
 \label{reheating electric power spectrum1}
\end{eqnarray}
respectively, with $r$ being the electromagnetic polarization index and runs from $r = \pm$, and furthermore 
$\theta^{(r)}_1 = \mathrm{Arg}[\alpha_r(k,\eta_f)]$ and $\theta^{(r)}_2 = \mathrm{Arg}[\beta_r(k,\eta_f)]$. 
Consequently, the total electromagnetic power spectrum is
\begin{eqnarray}
 \frac{\partial \rho_{em}}{\partial \ln{k}} = \frac{\partial \rho(\vec{B})}{\partial \ln{k}} + \frac{\partial \rho(\vec{E})}{\partial \ln{k}} 
 = \frac{1}{\pi^2}~\sum_{r=+,-}~\bigg(\frac{k^4}{a^4}\bigg)\bigg[\big|\alpha_r(k,\eta_f)\big|^2 + \big|\beta_r(k,\eta_f)\big|^2\bigg].
 \label{reheating em power spectrum}
\end{eqnarray}
 Furthermore, during the reheating epoch the helicity spectrum evolves as,
\begin{eqnarray}
P^{(h)}(k,\eta) &=&\bigg(\frac{k^3}{\pi^2a^3}\bigg)
\bigg\{\big|\beta_+ \big|^2 - \big|\beta_-\big|^2 
+ \sqrt{1 + \big|\beta_+\big|^2}~\big|\beta_+\big|~\cos{\big\{\theta^{(+)}_1 - \theta^{(+)}_2 - 2k(\eta - \eta_f)\big\}}\nonumber\\ 
&-&\sqrt{1 + \big|\beta_-\big|^2}~\big|\beta_-\big|~\cos{\big\{\theta^{(-)}_1 - \theta^{(-)}_2 - 2k(\eta - \eta_f)\big\}}\bigg\}~~,
\label{helicity5}
\end{eqnarray}
where we use $\big|\beta_{\pm}\big|^2 - \big|\alpha_{\pm}\big|^2 = 1$. 
It can be observed from the above equations that individual electric and magnetic spectrum depends on the phase factor 
evolution which is the well known Faraday effect. However,  the comoving total electromagnetic power spectrum is 
independent of time and behave as standard radiation field. Furthermore, due to zero electrical conductivity, the 
helicity spectrum also evolve with the phase factor during reheating.    
Moreover the relative phase factors between the Bogoliubov 
coefficients, from the numerical solutions of $A_{\pm}(k,\eta)$, are determined as,
\begin{eqnarray}
 \theta^{(+)}_1 - \theta^{(+)}_2&=&\mathrm{Arg}\big[\alpha_{+}(\eta_f)\beta_{+}^{*}(\eta_f)\big] \simeq 3.141\nonumber\\
 \theta^{(-)}_1 - \theta^{(-)}_2&=&\mathrm{Arg}\big[\alpha_{-}(\eta_f)\beta_{-}^{*}(\eta_f)\big] \simeq 3.141~~~.
 \label{relative phase factors}
\end{eqnarray}
 This can also be understood from the analytic solutions of $A_{\pm}(k,\eta)$ in the superhorizon limit. In particular we obtained 
$\lim_{|k\eta|\ll 1}A_{\pm} \sim (-k\eta)^{-m}$ which leads to $\lim_{|k\eta| \ll 1}A'_{\pm} \sim (-k\eta)^{-m-1}$. Therefore for $m = 1$ 
(i.e the parametric value for the scenario S2), $A'_{\pm}(k,\eta)$ dominates over the corresponding mode functions $A_{\pm}(k,\eta)$ at the limit 
$|k\eta| \rightarrow 0$ i.e towards the end of inflation. As a result,
\begin{eqnarray}
 \alpha_{\pm}(k,\eta_f) \simeq \frac{i}{\sqrt{2k}}~A_{\pm}'(k,\eta_f)~~~~~~~~~~,~~~~~~~~~~
 \beta_{\pm}(k,\eta_f) \simeq  \frac{-i}{\sqrt{2k}}~A_{\pm}'(k,\eta_f)~~~~.
 \label{reheating Bogoliubov5}
\end{eqnarray}
The above expressions, due to the presence of the $\pm i$ factor, immediately lead to the relative phase between the respective Bogoliubov coefficients as 
$\theta_1^{(\pm)} - \theta_2^{(\pm)} = \pi$, which 
is in agreement with the numerical estimation in Eq.(\ref{relative phase factors}).

Now we need to know the unknown part of the phase factor ``$\eta-\eta_f$'' which is defined as 
\begin{eqnarray}
 \eta - \eta_f = \int_{a_f}^{a}\frac{da}{a^2H}~~.
 \label{difference conformal time}
\end{eqnarray}
The presence of the Hubble parameter in Eq.(\ref{difference conformal time}) clearly indicates that in order to determine $\eta-\eta_f$ 
during the reheating phase, we need to state the background reheating dynamics, in particular how the inflaton energy density converts to other matter form. 
In the present context, we will consider two simple reheating mechanisms: (1) It is described by effective time independent equation of state and 
(2) perturbative reheating scenario. 

\subsection{Reheating with time independent equation of state: Evolution of Electromagnetic field}

In regard to the reheating process, as the first consideration, we follow the conventional reheating mechanism proposed 
by Kamionkowski et al. \cite{Dai:2014jja}, where the reheating phase is parametrized by a constant equation of state $\omega_\mathrm{eff}$. 
The reheating equation of state parameter (i.e $\omega_\mathrm{eff}$) remains constant during the entire reheating epoch and finally makes a 
sharp transition to the vale $1/3$ at the end of it, which in turn indicates the beginning of the radiation epoch. Due to the constant equation of state, 
the Hubble parameter during the reheating evolves as $H \propto a^{-\frac{3}{2}(1 + \omega_\mathrm{eff})}$. Moreover, 
the reheating e-fold number $N_\mathrm{re}$, and the reheating temperature ($T_\mathrm{re}$) 
can be expressed in terms of the $\omega_\mathrm{eff}$ and of some inflationary parameters by the following relations \cite{Dai:2014jja,Cook:2015vqa},
\begin{eqnarray}
N_{re}=\frac{4}{3\omega_\mathrm{eff}-1}\left[\frac{1}{4}ln(\rho_{f})-\frac{1}{4}ln\left(\frac{\pi^2 g_{re}}{30 }\right)
-\frac{1}{3}\left(\frac{43}{11 g_{s,re}}\right)- ln\left(\frac{a_0 T_0}{k}\right)- ln(H_{*}) + N_f \right]~~,
 \label{reheating e-folding}
\end{eqnarray}
\begin{eqnarray}
 T_{re}=\left(\frac{43}{11 g_s,re}\right)^{\frac{1}{3}}\bigg(\frac{a_0 T_0}{k}\bigg) H_* e^{-N_f}e^{-N_{re}},
 \label{reheating temperature}
\end{eqnarray}
where, recall, the subscript '*' denotes the value of a quantity at the horizon crossing instant of CMB scale and $N_f$ is the e-fold number 
of the inflation epoch. Moreover $\rho_f$ denotes the inflaton energy density at the end of inflation. The present CMB 
temperature is taken as $T_0 = 2.725\mathrm{K}$, $k \approx 0.05\mathrm{Mpc}^{-1}$ is the pivot scale and $a_0$ denotes the 
present time scale factor. The symbols $g_{s,re}$ and $g_{re}$ represent the degrees 
of freedom for entropy at reheating and the effective number of relativistic species upon 
thermalization respectively, and moreover they are taken to be the same, particularly $g_{s,re} = g_{re} \approx 100$.\\
It may be observed from the above equations that both $N_{re}$ and $T_{re}$ depend on the inflationary parameters, in particular on 
$H_*$ and $N_f$. In regard to the inflationary quantities, we first consider the scalar spectral index ($n_s$) and the scalar 
perturbation amplitude ($\mathcal{A}_s$), which, in the context of EFT, take the following forms \cite{Cheung:2007st}, 
\begin{eqnarray}
 n_s = 1 + \frac{4\dot{H}_*}{H_*^2}-\frac{\ddot{H}_*}{\dot{H}_*H_*}~~~~~~~~~,~~~~~~~~~
 \mathcal{A}_s = \frac{H_*^2}{4\epsilon_* M_{Pl}^2}
 \label{ns and As}
\end{eqnarray}
with $\epsilon_* = -\dot{H}_*/H_*^2$ and an overdot represents the differentiation with respect to the cosmic time. 
The first and second derivatives of $H_*$ (appearing in the expression of $n_s$) are connected to the first and second slow roll parameters respectively. 
To determine $N_f$, we expand the Hubble parameter around $t = t_*$ (i.e the horizon crossing instant of the CMB scale) as follows,
\begin{eqnarray}\label{inf_time1}
H(t) = H_* + \dot{H}_* (t-t_*) + \frac{1}{2} \ddot{H}_* (t-t_*)^2
\end{eqnarray}
where we retain the terms upto $\mathcal{O}(\ddot{H}_*)$, that means we consider upto the second slow roll parameter. If the duration of inflation 
is designated by $\Delta t$, then the Hubble parameter at the end of inflation ($H_f$) is given by,
\begin{eqnarray}\label{inf_time2}
H_{f} = H_* + \dot{H}_* \Delta t + \frac{1}{2}\ddot{H}_*\big(\Delta t\big)^2~~.
\end{eqnarray}
Inverting the above equation, we get the duration of inflation in terms of $H_*$, $H_f$, $\dot{H}_*$ and $\ddot{H}_*$ as,
\begin{eqnarray}\label{inf_time3}
\Delta t = \frac{|\dot{H}_*|}{\ddot{H}_*}\left(1-\sqrt{1-\frac{2 \ddot{H}_*}{|\dot{H}_*|^2}(H_* - H_{f})}\right)
\end{eqnarray}
with $|\dot{H}_*| = -\dot{H}_*$. Using the above expressions, we determine the inflationary e-folding number and is given by, 
\begin{eqnarray}\label{nk}
N_f&=&\int_{t_*}^{t_f}H(t)dt\nonumber\\
&=&\frac{H_* |\dot{H}_*|}{\ddot{H}_*}\left(1-\sqrt{1-\frac{2 \ddot{H}_*}{|\dot{H}_*|^2}(H_* - H_{f})}\right) 
- \frac{|\dot{H}_*|^3}{2 \ddot{H}_*^2}\left(1-\sqrt{1-\frac{2 \ddot{H}_*}{|\dot{H}_*|^2}(H_* - H_{f})}\right)^2\nonumber\\
&+&\frac{|\dot{H}_*|^3}{6 \ddot{H}_*^2}\left(1-\sqrt{1-\frac{2 \ddot{H}_*}{|\dot{H}_*|^2}(H_* - H_{f})}\right)^3
\end{eqnarray}
where, in the second line, we use the expression of $\Delta t$ from Eq.(\ref{inf_time3}). Furthermore, Eq.(\ref{ns and As}) leads to 
$\dot{H}_*$ and $\ddot{H}_*$ in terms of $n_s$, $\mathcal{A}_s$ as follows 
\begin{eqnarray}
 |\dot{H}_*| = \frac{H_*^4}{4\mathcal{A}_s M^2_{Pl}}~~~~~~~~~~,~~~~~~~~~~~~~
 \ddot{H}_* = \frac{H_*^5}{4\mathcal{A}_s M^2_{Pl}}\left(\frac{H_*^2}{\mathcal{A}_s M^2_{Pl}} - (1- n_s) \right)
 \label{1st and 2nd derivative of H}
\end{eqnarray}
respectively. Here we would like to mention that, as depicted in Fig.[\ref{plot_observable}], the observable quantities like scalar spectral index and 
tensor to scalar ratio are indeed compatible with the Planck constraints in the present EFT context where the EFT coefficients satisfy 
Eq.(\ref{EFT constraints modified2}). Thereby using the viable value(s) of $n_s$ and $\mathcal{A}_s$, 
in particular we will consider $n_s = 0.9649$ and $\ln{[10^{10}\mathcal{A}_s]} = 3.044$ (i.e the central values of $n_s$ and $\mathcal{A}_s$ in 
respect to the Planck constraints), one can determine $\dot{H}_*$ and $\ddot{H}_*$ from Eq.(\ref{1st and 2nd derivative of H}), which in turn will evaluate 
the inflationary e-fold number from Eq.(\ref{nk}).\\
Coming back to the quantity $\eta-\eta_f$, we first consider the behaviour of the Hubble parameter during the reheating stage. Due to the fact that 
the effective EoS remains constant, 
the Hubble parameter behaves as $H \propto a^{-\frac{3}{2}(1 + \omega_\mathrm{eff})}$, and as a result, the term 
$\eta - \eta_f$ boils down to the following simple form:
\begin{eqnarray}
 \eta - \eta_f = \frac{2}{\big(1 + 3\omega_\mathrm{eff}\big)}\bigg[\frac{1}{aH} - \frac{1}{a_fH_f}\bigg],
 \label{Kamionkowski special term}
\end{eqnarray}
where we use Eq.(\ref{difference conformal time}). Due to the above expression of $\eta-\eta_f$ along with the fact that 
$\big|\beta_{\pm}(\eta_f)\big|^2 \gg 1$, the magnetic power spectrum during the reheating era becomes \cite{Kobayashi:2019uqs}
\begin{eqnarray}
 \frac{\partial \rho(\vec{B})}{\partial \ln{k}} = \frac{1}{\pi^2}\bigg(\frac{k^4}{a^4}\bigg)~\sum_{r=+,-}~\big|\beta_r\big|^2~ 
 \bigg\{\mathrm{Arg}\big[\alpha_r~\beta_r^{*}\big] - \pi - \bigg(\frac{4k}{3\omega_\mathrm{eff} + 1}\bigg)
 \bigg(\frac{1}{aH} - \frac{1}{a_fH_f}\bigg)\bigg\}^2.
 \label{reheating magnetic power spectrum2}
\end{eqnarray}
The above equation clearly indicates that the magnetic power evolution during reheating scales by two different terms: (1) the conventinal one 
that goes as $1/a^4$ and (2) the term that redshifts as $\propto a^{-6}H^{-2}$, emerging from $\frac{1}{aH}$ in the right hand side of 
Eq.(\ref{reheating magnetic power spectrum2}). Due to the reason that the universe undergoes through deceleration, the EoS follows 
$\omega_\mathrm{eff} > -\frac{1}{3}$, and thus the magnetic power effectively scales by the component $\propto a^{-6}H^{-2}$. 
In the context of instantaneous reheating, the magnetic power evolves as $a^{-4}$ from end of inflation until today, where as 
if one considers intermediate reheating phase with negligible conductivity the evolution of magnetic power will be as described above. 
Due to this fact present strength of the magnetic field can be lager compared to that for the instantaneous reheating case. 
This particular phenomena can make S2-like magnetogenesis scenario viable which we will elaborate next. 
The electric power in the reheating epoch goes evolves as,
\begin{eqnarray}
 \frac{\partial \rho(\vec{E})}{\partial \ln{k}} = \frac{1}{\pi^2}\bigg(\frac{k^4}{a^4}\bigg)~\sum_{r=+,-}~\big|\beta_r(\eta_f)\big|^2
 \label{reheating electric power spectrum2}
\end{eqnarray}
Hence the electric power evolve exactly like radiation during reheating at large scale. Finally the behaviour of the helicity 
spectrum takes the following form in the large $|\beta(k,\eta_f)|$ limit,
\begin{eqnarray}
P^{(h)} = \bigg(\frac{k^3}{\pi^2 a^3}\bigg)\bigg\{\big|\beta_+(k,\eta_f)\big|^2 - \big|\beta_-(k,\eta_f)\big|^2\bigg\} 
\bigg\{\frac{4k}{3\omega_\mathrm{eff} + 1}\bigg(\frac{1}{aH} - \frac{1}{a_fH_f}\bigg)\bigg\}^2 \propto \frac{1}{a^5 H^2} + \cdots~~.
\label{helicity7}
\end{eqnarray}
During reheating the Hubble parameters behaves as  $H = a^{-\frac{3}{2}(1 + \omega_\mathrm{eff})}$.  Hence, from Eq.(\ref{helicity7}), 
the leading behaviour of the helicity density will be, $a^3 P^{(h)} \sim a^{1+3\omega_\mathrm{eff}}$. 
For all reheating equation of state $\omega_\mathrm{eff} > -\frac{1}{3}$, 
the comoving helicity during the reheating era monotonically grows with time.  

\subsubsection{Present day magnetic field and constraints on reheating parameters, $\omega_{eff}$}

Here we may recall that the scenario S2 with an instantaneous reheating mechanism does not predict a sufficient magnetic strength in the 
present epoch, in particular it predicts $B_0 = 10^{-40}\mathrm{G}$ 
as shown in Eq.(\ref{magnetic strength 4-II}).
Given the characteristic difference in evolution of magnetic power spectrum during reheating, we see how those models can be made viable with 
the CMB observations. With this motivation in mind we are now in position to constrain such model with respect to the observation. 
For this let us first express the Hubble parameter at the end of reheating ($H_{re}$) in terms of that of the end of inflation ($H_f$) as
\begin{eqnarray}
H_{re} = H_f\bigg(\frac{a_{re}}{a_f}\bigg)^{-\frac{3}{2}(1 + \omega_{eff})},
\label{reheating end Hubble parameter}
\end{eqnarray}
where the suffix '\textit{re}' denotes the end point of reheating and  the scale factor $a_{re}$ can be identified as 
${a_{re}}/{a_f} = e^{N_\mathrm{re}}$ with $N_\mathrm{re}$ being the e-fold number of the reheating epoch and given in Eq.(\ref{reheating e-folding}). 
Consequently, from Eq.(\ref{reheating magnetic power spectrum2}) we evaluate the magnetic power spectrum at the end instant of reheating, as
\begin{eqnarray}
\frac{\partial \rho(\vec{B})}{\partial \ln{k}}\bigg|_{re} = \bigg(\frac{k^4}{a_{re}^4 \pi^2}\bigg)\sum_{r=\pm}~\big|\beta_r \big|^2~ 
\bigg\{\mathrm{Arg}\big[\alpha_r\beta_r^{*} \big] - \pi - \bigg(\frac{4k}{(3\omega_\mathrm{eff} + 1)a_fH_f}\bigg)
\bigg[\bigg(\frac{H_f}{H_{re}}\bigg)^{\delta} - 1\bigg]\bigg\}^2~~~,
\label{reheating magnetic power spectrum4}
\end{eqnarray}
where, $\delta = (3\omega_\mathrm{eff} + 1)/(3\omega_\mathrm{eff}+3)$.
After the reheating era the conductivity of the universe is assumed to be sufficiently large, so that, the electric field dies out very fast. 
Therefore, Faraday induction will halt. Consequently the magnetic field begins to evolve according to radiation as $a^{-4}$ till today. 
The present-day magnetic power spectrum then becomes, 
\begin{eqnarray}
\frac{\partial \rho(\vec{B})}{\partial \ln{k}}\bigg|_{0} = \bigg(\frac{a_{re}}{a_0}\bigg)^4~\frac{\partial \rho(\vec{B})}{\partial \ln{k}}\bigg|_{{re}}
\label{reheating magnetic strength K1}
\end{eqnarray}
and, as a result, Eq.(\ref{reheating magnetic power spectrum4})  leads to the current magnetic strength, as
\begin{eqnarray}
B_0 = \frac{\sqrt{2}}{\pi}\bigg(\frac{k}{a_0}\bigg)^2\sqrt{\sum_{r=\pm}\big|\beta_r\big|^2~ 
	\bigg\{\mathrm{Arg}\big[\alpha_r \beta_r^{*} \big] - \pi - \bigg(\frac{4}{3\omega_\mathrm{eff} + 1}\bigg)
	\bigg(\frac{k}{a_fH_f}\bigg)\bigg[\bigg(\frac{H_f}{H_{re}}\bigg)^{\delta} - 1\bigg]\bigg\}^2}~~~,
\label{reheating magnetic strength K2}
\end{eqnarray}
The above expression clearly indicates that the magnetic field's current amplitude explicitly 
depends on the reheating parameters $(\omega_\mathrm{eff}, H_{re}$), and inflationary $n_s$. As a consequence, probing 
$B_0$ and its nature opens up a new window to look into the reheating phase. Given an inflationary model, 
the expression in Eq.(\ref{reheating magnetic strength K2}) establishes the connection 
between the current magnetic strength $B_0$ and the reheating EoS parameter $\omega_\mathrm{eff}$. 
Thus we may argue that the effective equation of state is no longer a free parameter, rather it is fixed by the $B_0$ via CMB.\\

\begin{figure}[t]
	\begin{center}
		\centering
		\includegraphics[width=3.0in,height=2.2in]{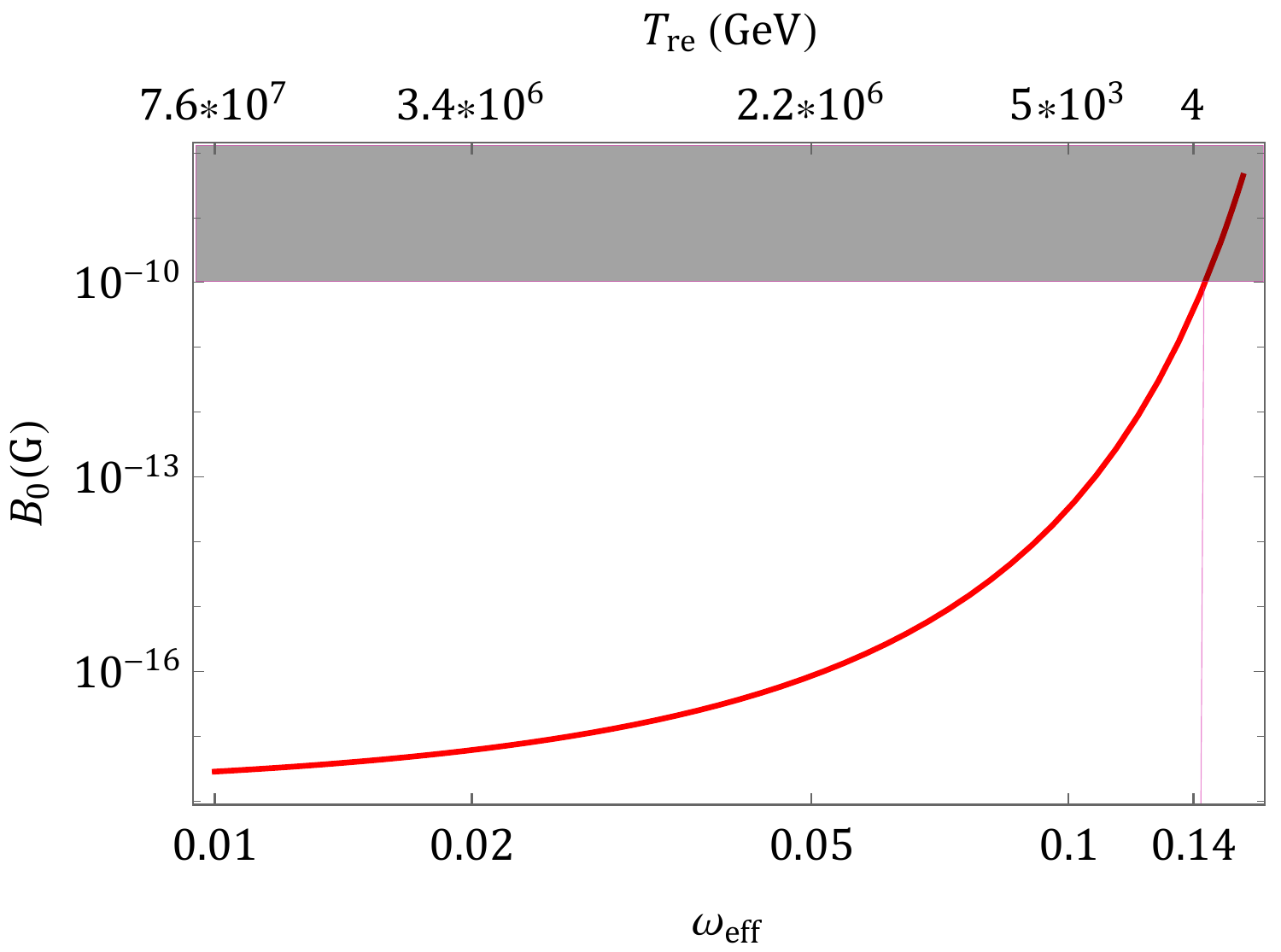}
		\includegraphics[width=3.0in,height=2.2in]{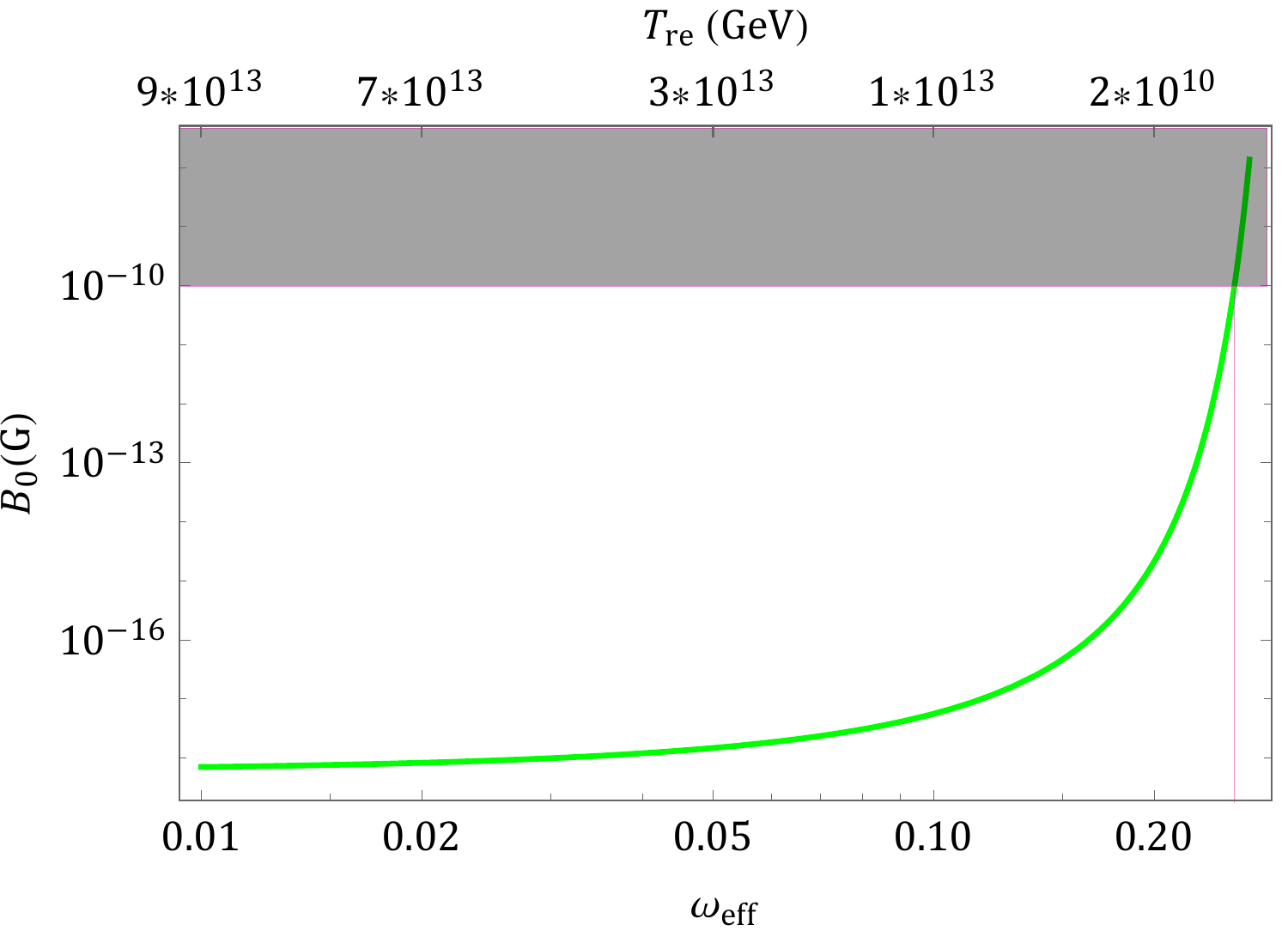}
		\caption{$B_0$ (in Gauss units) vs $\omega_\mathrm{eff}$ with $k = 0.05\mathrm{Mpc}^{-1}$ in the Kamionkowski like reheating scenario. 
			The reheating temperature for different values of $\omega_\mathrm{eff}$ is shown in the upper label of the $x$-axis. In the left plot: 
			$H_* = 10^{-5}M_{Pl}$, $n_s = 0.9649$, $\ln{\big[10^{10}\mathcal{A}_s\big] = 3.044}$, $N_f = 50$ and in the right plot: 
			$H_* = 10^{-5}M_{Pl}$, $n_s = 0.9649$, $\ln{\big[10^{10}\mathcal{A}_s\big] = 3.044}$, $N_f = 55$.}
		\label{plot_magnetic strength K}
	\end{center}
\end{figure}

Having obtained the final expression of $B_0$, now we confront our model with the CMB observations which put a constraint 
on the current magnetic strength, as $10^{-10}\mathrm{G} \lesssim B_0 \lesssim 10^{-22}\mathrm{G}$ 
\cite{Widrow:2002ud,Kandus:2010nw,Durrer:2013pga}. For this purpose, we need ${H_f}/{H_{re}}$, which depends on 
$N_\mathrm{re}$, as discussed after Eq.(\ref{reheating end Hubble parameter}). 
In our effective theory framework, we have three main input parameters, the inflationary energy scale 
$H^*$ and the inflationary e-folding number $N_f$, and the reheating equation of state $\omega_{eff}$. 
All the other parameters can be computed such as $N_\mathrm{re}, H_f, \dot{H}_*$ and $\ddot{H}_*$.
 Given the observed central values $n_s = 0.9649$, $\ln{\big[10^{10}\mathcal{A}_s\big] = 3.044}$, for our estimation 
 we consider two different sets of values for   
$(H_*, N_f)$  particular, we consider - (1) $(H_*, N_f) = (10^{-5}M_{Pl}, 50)$ and (2) $(H_*, N_f) = (10^{-5}M_{Pl}, 55)$. 
The first and second sets correspond to the Hubble parameter at the end of inflation as 
$H_f = 3.61\times10^{-6}M_{Pl}$ and $H_f = 1.62\times10^{-6}M_{Pl}$ respectively, i.e $H_f$ decreases with increasing 
e-folding number, as expected. With such considerations, we plot $B_0$ versus $\omega_\mathrm{eff}$ by using Eq.(\ref{reheating magnetic strength K2}) 
in Fig.\ref{plot_magnetic strength K}.
It clearly demonstrates that the theoretical prediction of $B_0$ lies well within the observational constraints 
for a certain range of values of the reheating EoS parameter, given by : $0.01 \lesssim \omega_\mathrm{eff} \lesssim 0.14$ for $N_f = 50$ and 
$0.01 \lesssim \omega_\mathrm{eff} \lesssim 0.25$ for $N_f = 55$, which are also consistent with BBN constraint 
on the minimum reheating temperature $\sim 10^{-2}\mathrm{GeV}$. 

\begin{figure}[t]
\begin{center}
 \centering
 \includegraphics[width=3.5in,height=2.5in]{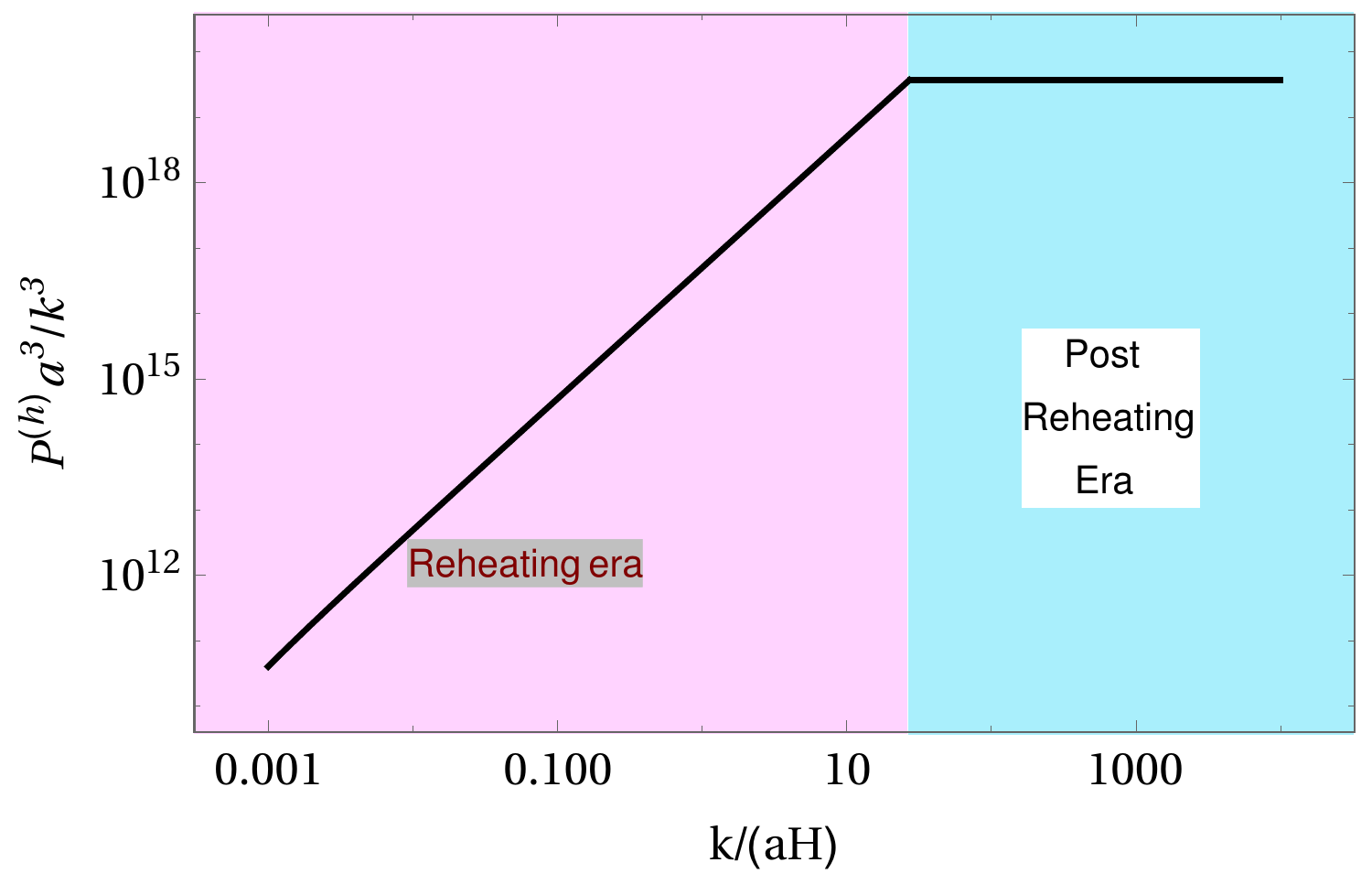}
 \caption{$P^{(h)}a^3/k^3$ versus $k/(aH)$ during and after reheating era. The reheating EoS is taken as $\omega_\mathrm{eff} = 0$ for which 
 $\frac{k}{aH} \propto k\sqrt{a} \propto k\eta$.}
 \label{plot_helicity_reheating}
\end{center}
\end{figure}

Finally let us also point out the how the helical nature of the magnetic field spectrum evolves which is shown in the 
Fig.[\ref{plot_helicity_reheating}]. The figure depicts that the comoving helicity density grows with time during the reheating stage. 
However, during the post reheating phase, the universe is considered to be a very good conductor, 
which in turn makes the electromagnetic mode functions 
constant over time. As a result, the helicity goes as $1/a^3$ or equivalently the comoving helicity 
becomes constant during the after reheating phase to the present epoch. Important to realize that introduction of reheating has increased the 
comoving helicity density by manifold compared to that the helicity density produced for the instantaneous reheating case. 

\subsection{Perturbative reheating with time dependent EoS: Evolution of magnetic field}

In this section we consider perturbative reheating scenario \cite{Maity:2018dgy}, where the inflaton energy density ($\rho_\mathrm{\phi}$)  
decays into the radiation energy density ($\rho_\mathrm{R}$) governed by the Boltzmann equations 
Eqs.(\ref{Boltzmann equation1}) and (\ref{Boltzmann equation2}). This essentially leads to the 
time dependent effective equation of state (EoS) during phase, which is defined by
\begin{eqnarray}
 \omega_\mathrm{eff} = \frac{3\omega_{\phi}\rho_{\phi} + \rho_{R}}{3\big(\rho_{\phi} + \rho_{R}\big)} .
\label{effective EoS}
\end{eqnarray}
Where, $\omega_{\phi}$  symbolizes the 
time averaged EoS of the inflaton field. For example during reheating, for oscillating canonical inflaton with potential $V(\phi) \propto \phi^p$, 
assumes the time averaged EoS, 
$\omega_\mathrm{\phi} = {(p-2)}/{(p+2)}$. However, effective equation of state $\omega_{eff}$ naturally  \cite{Maity:2018dgy} depend on time. 
Initially, $\omega_\mathrm{eff}$ behaves as approximately equal to $\omega_{\phi}$, and after a certain instance of time 
it smoothly transits to the value $\approx 1/3$ setting the  beginning of the radiation epoch. This 
scenario is different from that of the earlier one where 
the inflaton energy density is assumed to be instantaneously converted into radiation energy density at the end of reheating.

\begin{figure}[t!]
\centering
\includegraphics[width=9.50cm]{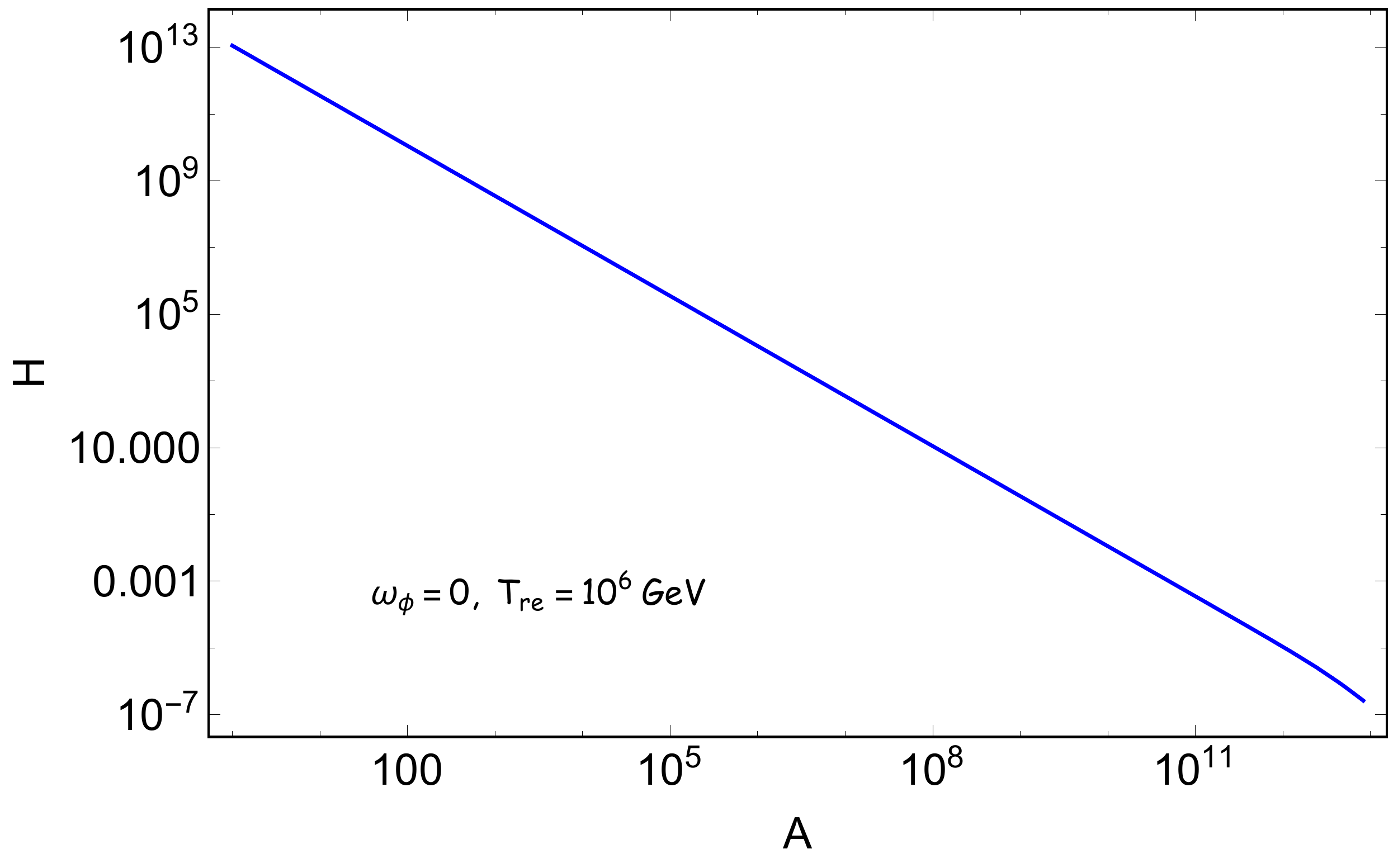}
\caption{The background Hubble parameter $H(\xi)$ versus $\xi$ during the reheating era i.e for $1 \leq \xi \leq \xi_{re}$. The plot corresponds to 
$\omega_{\phi} = 0$ and $T_\mathrm{re} = 10^{6}\mathrm{GeV}$.} 
\label{plot_solution_reheating}
\end{figure}

The set of Boltzmann equations governing the evolution of $\rho_{\phi}$ and $\rho_{R}$ are given by,
\begin{eqnarray}
 \frac{d\Phi}{d\xi} +\frac{\sqrt{3}M_{Pl} \Gamma_{\phi}}{H^2_{f}}(1+\omega_{\phi})\frac{\xi^{1/2}\Phi}{\Phi \xi^{-3\omega_{\phi}} + R \xi^{-1}} = 0
 \label{Boltzmann equation1}
\end{eqnarray}
and
\begin{eqnarray}
 \frac{dR}{d\xi} - \frac{\sqrt{3}M_{Pl} \Gamma_{\phi}}{H^2_{f}}(1+\omega_{\phi})\frac{\xi^{\frac{3(1- 2\omega_{\phi}}{2}} \Phi}
 {\Phi \xi^{-3 \omega_{\phi}} + R \xi^{-1}} = 0~~~,
 \label{Boltzmann equation2}
\end{eqnarray}
respectively. The Boltzmann equations are expressed in terms of a rescaled scale factor $\xi(a) = a/a_f$ for convenience, with $a_f$ being the scale factor 
at the end of inflation. Furthermore 
$\Gamma_\mathrm{\phi}$ symbolizes the decay rate from inflaton to radiation energy density 
and the comoving densities are rescaled in terms of the dimensionless variable as, 
\begin{eqnarray}\label{rescale}
\Phi = \bigg(\frac{\rho_{\phi}}{H_\mathrm{f}^4}\bigg) \xi^{3\big(1 + \omega_{\phi}\big)}~~~~~~;~~~~~~~~R = \bigg(\frac{\rho_R}{H_\mathrm{f}^4}\bigg) \xi^4~~.
 \end{eqnarray}
For solving the above Boltzmann equations, the natural initial conditions will be set at the end of inflation as follows,
 \begin{equation}
 \Phi(\xi = 1) = \rho_f/H_f^4~~~~~~\mathrm{and}~~~~~~R(\xi = 1) = 0~,
 \label{boundary Boltzmann equation1}
\end{equation}
$\rho_f$ is the inflaton energy density at $\xi = 1$. 
Furthermore, the reheating temperature is identified from the radiation temperature $T_\mathrm{rad}$ at the point of 
$H(t_\mathrm{re}) = \Gamma_\mathrm{\phi}$ ($t_\mathrm{re}$ denotes the end of reheating), 
when maximum inflaton energy density transfer into radiation.
\begin{eqnarray} \label{reheating 2}
T_{re}= T_{rad}^{end}= \left(\frac{30}{\pi^2 g_{re}}\right)^{1/4}\big[\rho_{R}(\Gamma_\phi,\xi_{re},n_{s})\big]^{1/4}~.
\end{eqnarray}
with $\xi_{re}$ being the rescaled factor at the end of reheating i.e $\xi_{re} = a_{re}/a_f$. 
From the entropy conservation of thermal radiation, the relation among $T_\mathrm{rad} = T_\mathrm{re}$ at equilibrium, 
and $(T_0, T_{\nu 0} = (4/11)^{1/3} T_0)$, 
temperature of the CMB photon and neutrino background at the present day respectively, can be written as 
\begin{eqnarray}\label{entropy}
g_{re} T_{re}^3 = \left(\frac {a_0}{a_{re}}\right)^3\left( 2 T_0^3 + 6 \frac 7 8 T_{\nu 0}^3\right).
\end{eqnarray}
Using above equation, one arrives at the following well known relation
\begin{eqnarray}\label{eqtre}
T_{re}= \left(\frac{43}{11 g_{re}}\right)^{\frac 1 3}\left(\frac{a_0T_0}{k}\right) H_* e^{-N_\mathrm{f}} e^{-N_\mathrm{re}}~~,
\end{eqnarray}
where $N_{re}$ is the e-folding number of the reheating era. To establish one to one correspondence between $T_\mathrm{re}$ and $\Gamma_\mathrm{\phi}$, 
we combine the equations (\ref{reheating 2}) and (\ref{eqtre}). To fixed the values of decay width $\Gamma_\mathrm{\phi}$ in terms of 
spectral index ($n_s$), we use one further condition at the end of the reheating
\begin{eqnarray} \label{reheating 1}
 H^2(\xi_{re}) = \frac{1}{\xi_\mathrm{re}^2}\bigg(\frac{d\xi_\mathrm{re}}{dt}\bigg)^2 
 = \frac{\rho_\mathrm{\phi}(\Gamma_\mathrm{\phi},\xi_\mathrm{re},n_{s})+ \rho_{R}(\Gamma_\phi,\xi_{re},n_{s}))}{3 M_\mathrm{Pl}^2} 
 = \Gamma_\mathrm{\phi}^2~~.
\end{eqnarray}
With the above equipments, we solve the Hubble parameter ($H^2(\xi) = \frac{1}{3M_{Pl}^2}\big(\rho_{\phi} + \rho_R\big)$) numerically along with  the coupled Boltzmann differential equations (\ref{Boltzmann equation1}) and (\ref{Boltzmann equation2}). The numerical solutions of $H(\xi)$ 
is depicted in Fig.[\ref{plot_solution_reheating}] which corresponds to a particular set: 
$T_\mathrm{re} = 10^{6}\mathrm{GeV}$ and $\omega_{\phi} = 0$ respectively. 

In the perturbative reheating scenario, the quantity $\eta - \eta_f$ of Eq.(\ref{difference conformal time}) can be expressed as,
\begin{eqnarray}
 \eta - \eta_f = \bigg(\frac{1}{a_fH_\mathrm{f}}\bigg)\int_{1}^{\xi}\frac{d\xi}{\xi^2}\frac{\sqrt{3}M_{Pl}H_\mathrm{f}}{\sqrt{\rho_r (\xi)+ \rho_{\phi}(\xi) }}~~,
 \label{perturbative special term}
\end{eqnarray}
where $\rho_{\phi}$ and $\rho_R$ are governed by the coupled Eqs.(\ref{Boltzmann equation1}) and (\ref{Boltzmann equation2}). Due to the complicated nature 
of the Boltzmann equations, the integral in Eq.(\ref{perturbative special term}) may not be performed in a closed form, however we will perform 
it numerically for various values of $\omega_{\phi}$. The above expression of $\eta-\eta_f$ immediately leads to the magnetic power spectrum 
during the reheating era as follows,
\begin{eqnarray}
 \frac{\partial \rho(\vec{B})}{\partial \ln{k}} = \frac{1}{\pi^2}\bigg(\frac{k^4}{a^4}\bigg)~\sum_{r=\pm}~\big|\beta_r\big|^2~ 
 \bigg\{\mathrm{Arg}\big[\alpha_r \beta_r^{*}\big] - \pi - 
 2\bigg(\frac{k}{a_fH_\mathrm{f}}\bigg)\int_{1}^{\xi}\frac{d\xi}{\xi^2}\frac{\sqrt{3}M_{Pl}H_\mathrm{f}}{\sqrt{\rho_r (\xi)+ \rho_{\phi}(\xi) }}\bigg\}^2.
 \label{reheating magnetic power spectrum3}
\end{eqnarray}
 Furthermore, due to the fact that 
$\big|\beta_r(\eta_f)\big| \gg 1^2$, the electric power spectrum during the reheating phase goes as,
\begin{eqnarray}
 \frac{\partial \rho(\vec{E})}{\partial \ln{k}} = \frac{1}{\pi^2}\bigg(\frac{k^4}{a^4}\bigg)~\sum_{r=\pm}~\big|\beta_r(\eta_f)\big|^2~~.
 \label{reheating electric power spectrum3}
\end{eqnarray}
The above expressions will be used in investigating whether the scenario S2 with a reheating phase, characterized by perturbative mechanism, 
will predict sufficient magnetic strength at present universe. We will analyze this for different values of $\omega_{\phi}$ in Sec.[\ref{sec_present magnetic 
strength perturbative}], which in turn will help to probe various informations of the reheating phase.

\subsubsection{Present day magnetic field and constraints on perturbative reheating}\label{sec_present magnetic strength perturbative}
\begin{figure}[t]
\begin{center}
\centering
\includegraphics[width=3.0in,height=2.2in]{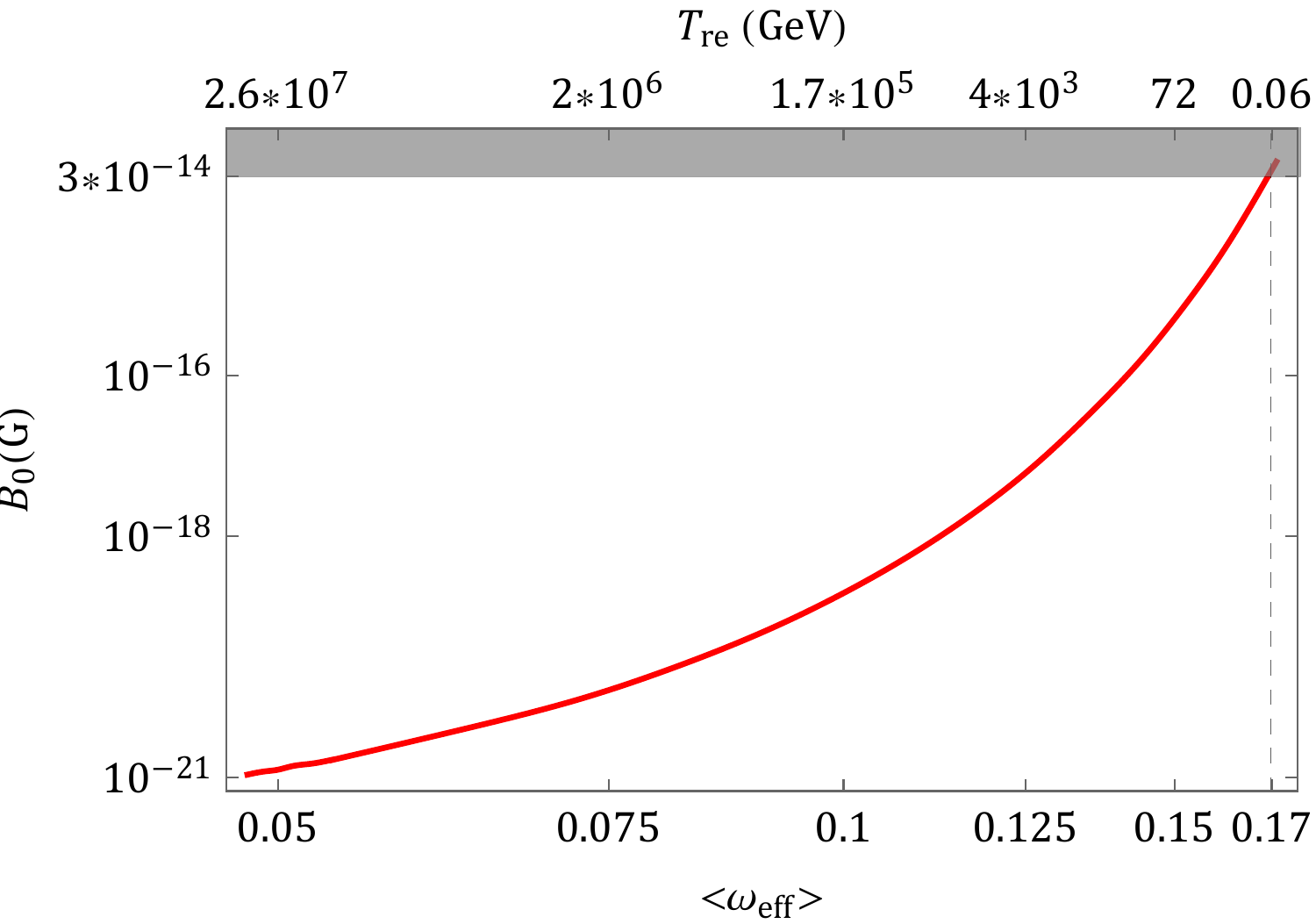}
\includegraphics[width=3.0in,height=2.2in]{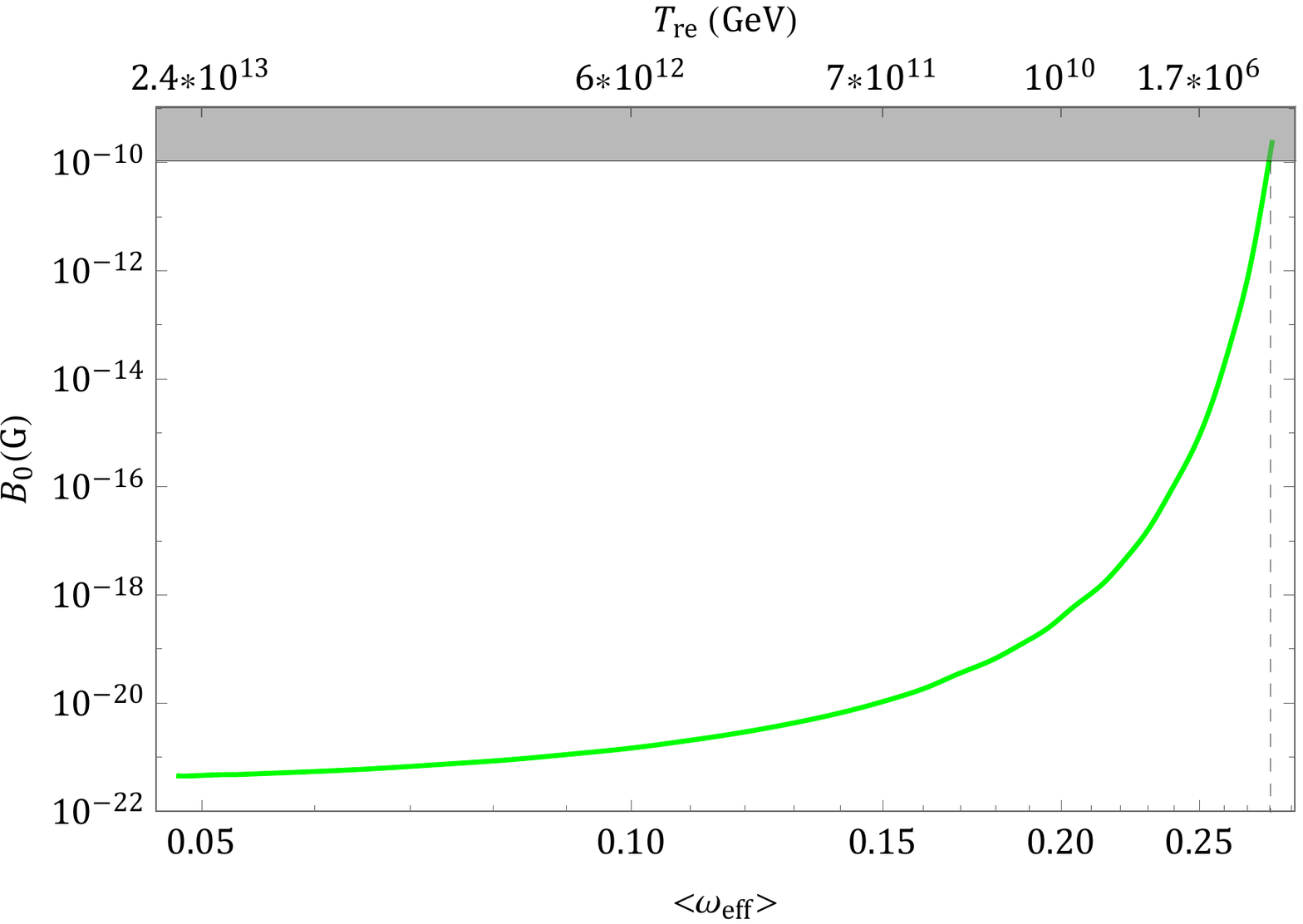}
\caption{$B_0$ (in Gauss units) vs $\langle\omega_\mathrm{eff}\rangle$ with $k = 0.05\mathrm{Mpc}^{-1}$ in the perturbative reheating scenario. 
The reheating temperature for different values of $\langle\omega_\mathrm{eff}\rangle$ is shown in the upper label of the $x$-axis. In the left plot: 
$H_* = 10^{-5}M_{Pl}$, $n_s = 0.9649$, $\ln{\big[10^{10}\mathcal{A}_s\big] = 3.044}$, $N_f = 50$ and in the right plot: 
$H_* = 10^{-5}M_{Pl}$, $n_s = 0.9649$, $\ln{\big[10^{10}\mathcal{A}_s\big] = 3.044}$, $N_f = 55$.} 
 \label{plot_magnetic strength_pert}
\end{center}
\end{figure}

Following this same methodology considered for the earlier reheating scenario, we will see hos perturbative reheating dynamics is 
constrained by the present value of the magnetic field. During the perturbative reheating the effective EoS $\omega_\mathrm{eff}$ time dependent.  
However, to set constrain, we may introduce an average effective EoS defined by,
\begin{eqnarray}
 \langle{\omega}_\mathrm{eff}\rangle = \frac{1}{\xi_{re}-1}\int_1^{\xi_{re}}\omega_\mathrm{eff}~d\xi = 
 \frac{1}{\xi_{re}-1}\int_1^{\xi_{re}}\bigg\{\frac{3\omega_{\phi}\rho_{\phi} + \rho_{R}}{3\big(\rho_{\phi} + \rho_{R}\big)}\bigg\}d\xi
 \label{average effective EoS}
\end{eqnarray}
where we use Eq.(\ref{effective EoS}) and recall, $\xi = 1$ to $\xi = \xi_{re}$ denotes the reheating phase.  
It may be observed from Eq.(\ref{average effective EoS}) that $\langle{\omega}_\mathrm{eff}\rangle$ contains 
the information of background evolution of $\rho_{\phi}$ and $\rho_R$, and also depends on the inflaton EoS ($\omega_{\phi}$). 
Following Eq.(\ref{reheating magnetic power spectrum3}), and the detailed procedure discussed in the previous section, 
one obtains the magnetic field strength at present epoch as,
\begin{eqnarray}\label{magnetic strength per2}
 B_0 = \frac{\sqrt{2}}{\pi}\bigg(\frac{k}{a_0}\bigg)^2\sqrt{\sum_{r=\pm}\big|\beta_r\big|^2~ 
 \bigg\{\mathrm{Arg}\big[\alpha_r~\beta_r^{*}\big] - \pi - 
 \bigg(\frac{2k M_{pl}}{a_fH_\mathrm{f}}\bigg)\int_{1}^{\xi_{re}}\frac{d\xi}{\xi^2}\frac{\sqrt{3}H_\mathrm{f}}
 {\sqrt{\rho_r (\xi)+ \rho_{\phi}(\xi) }}\bigg\}^2}
\end{eqnarray}
For this purpose, we will take same set of parameter values as before for our numerical computation, 
$(H_*, N_f) = (10^{-5}M_{Pl}, 50)$ and (2) $(H_*, N_f) = (10^{-5}M_{Pl}, 55)$. Given those values we first need to evaluate 
$\xi_{re}$ appearing in the upper limit of the integral, which is done by simultaneously satisfying the following equations:
\begin{eqnarray}\label{upper limit}
H(\xi_{re}) = \Gamma_{\phi}~~,~~
T_{re} = \left(\frac{30}{\pi^2 g_{re}}\right)^{1/4}\rho_R^{1/4}~~,~~
T_{re} = \left(\frac{43}{11 g_s,re}\right)^{\frac{1}{3}}\frac{a_0 T_0}{k} H_* e^{-N_f}e^{-N_{re}}
\end{eqnarray}
Using $\xi_{re}$ and the numerical solutions of $\rho_{\phi}$ and $\rho_R$ from the Boltzmann equations, 
we get the variation of $B_0$ with $\omega_{\phi}$, which is depicted in Fig.\ref{plot_magnetic strength_pert}. 
It is with respect to the average effective EoS of the reheating era, i.e in respect to 
$\langle\omega_\mathrm{eff}\rangle$ defined in Eq.(\ref{average effective EoS}).
As we may observe that similar to the earlier reheating case, the current magnetic strength 
in the perturbative reheating mechanism increases with the average effective EoS 
(i.e $\langle\omega_\mathrm{eff}\rangle$, defined in Eq.(\ref{average effective EoS})) and lies within the observational constraints 
for suitable regime of reheating parameters. However the viable range of the reheating parameters differ in the two reheating mechanisms 
respectively. In particular, the compatibility of the scenario S2, with a reheating phase characterized by perturbative mechanism, 
both with the CMB observations on $B_0$ and with the BBN constraint on the reheating temperature leads to a constraint on 
$\langle\omega_\mathrm{eff}\rangle$ as: $0.05 \lesssim \langle\omega_\mathrm{eff}\rangle \lesssim 0.17$ for $N_f = 50$ and 
$0.04 \lesssim \langle\omega_\mathrm{eff}\rangle \lesssim 0.27$ for $N_f = 55$ respectively.

\section{Conclusion}
Effective field theory is a powerful tool in various branches of physics. In the present work we studied in detail inflationary 
magnetogenesis in this EFT framework. In the cosmological universe, four dimensional diffeomorphism symmetry is generically broken down to 
spatial diffeomorphism. Using this remaining symmetry we have written down most general action upto quadratic order in scalar, tensor, 
electromagnetic vector fluctuation. In this framework, electromagnetic sector automatically breaks conformal invariance 
which plays the crucial role in producing the gauge field from the quantum vacuum. We have also considered the parity broken term which further 
gives rise the helical magnetic has great observational significance. 
The form of the coupling functions have been considered in such a way that during inflation electromagnetic field evolves non-trivially, 
however, at late times (in particular after the end of inflation), 
the conformal symmetry of the EM field is restored and consequently the standard Maxwellian evolution is restored.

We have explored the evolution of the EM, scalar and tensor perturbation, and determine the primordial power spectrum taking 
into account PLANCK, and Large scale magnetic field observation. In regard to the cosmological 
evolution of the electric and magnetic fields, we discussed possibilities of scale invariant magnetic or electric 
power spectrum particularly at the superhorizon scale.  
If we consider instantaneous reheating after the inflation,
it is observed that the scale invariant magnetic field scenario is observationally compatible, however, suffers from the 
backreaction problem. On the other hand, scale invariant electric field scenario  
is indeed free from the backreaction problem, but does not predict sufficient amount of magnetic strength in the present universe. 

To cure this problem, we next introduce reheating phase with non-trivial evolution dynamics
with non-zero e-folding number. Two different reheating mechanisms are considered: 
(i) Reheating dynamics is  \cite{Dai:2014jja} governed by a time independent effective equation of state ($\omega_\mathrm{eff}$). 
(ii) As a second possibility, we consider perturbative reheating scenario where effective equation of state is time evolving due to 
non-trivial decay of inflaton into the radiation. 
Because of reheating phase post-inflation dynamics of EM field becomes non-trivial. 
Most interesting case arises when one considers vanishing electrical conductivity during this phase, 
which induces additional magnetic field from  non-vanishing electric field due to well known Faraday effect. 
However, after the end of reheating, this effect ceases to exist as universe becomes  good conductor and, 
hence, the electric field vanishes. Specifically, the magnetic field energy density evolves as $a^{-6}H^{-2}$ during the reheating era, and 
after the reheating, as $a^{-4}$. 
Furthermore, we observed that reheating also helps increase the helicity of the magnetic field if 
parity violation is introduced in the system as shown in the Fig.\ref{plot_helicity_reheating}.
Reheating phase, therefore, helps enhancing the present value of  
magnetic field as compared to ordinary instantaneous reheating case.
Our detail analysis shows that this is precisely the mechanism which can provide right magnitude of 
the present magnetic field for the scale invariant electric field scenario. However, this mechanism does not help to 
cure the problem for the scale invariant magnetic case. 
The reason being that the backreaction problem occurs during the inflation and thus can not be rescued by any post inflationary phase. 
We will discuss the possible resolution of this problem in our forthcoming paper.

Most importantly, introducing reheating era not only helps to obtain the right magnitude of the large scale magnetic field, 
it allows one to obtain valuable information about the 
reheating EoS parameter ($\omega_\mathrm{eff}$) which in turn can potentially constraint the inflationary model itself, 
which has been discussed before in some of our recent works \cite{Bamba:2020qdj,Haque:2020bip}. 
Therefore, probing large scale magnetic field opens up a new probe to look into the reheating phase which is otherwise very difficult to constrain. 
Combining CMB, presently observed constraint on $B_0$ and the BBN constraint, our analysis restrict the value of  
of $\omega_\mathrm{eff}$ as follows: (i) $0.01 \lesssim \omega_\mathrm{eff} \lesssim 0.14$ for $N_f = 50$ and 
$0.01 \lesssim \omega_\mathrm{eff} \lesssim 0.25$ for $N_f = 55$ for the reheating scenario where EoS is constant.  and (ii) 
$0.05 \lesssim \langle\omega_\mathrm{eff}\rangle \lesssim 0.17$ for $N_f = 50$ and 
$0.04 \lesssim \langle\omega_\mathrm{eff}\rangle \lesssim 0.27$ for $N_f = 55$ for the perturbative reheating scenario 
(recall, $N_f$ is the inflationary e-folding number and 
$\langle\omega_\mathrm{eff}\rangle$ is the average effective EoS during the reheating era defined in Eq.(\ref{average effective EoS})). 
This provides a viable constraint on the reheating EoS parameter from CMB observations.

\appendix
\section{Discussion on several magnetogenesis models and their equivalence with the EFT formalism}\label{appendix}
As we have already mentioned in Sec.\ref{sec_model} 
that in this magnetogenesis scenario we can find one to one correspondence between the EFT approach and several well established 
models of inflationary magnetogenesis, here in this section we discuss that various magnetogenesis models can be 
embedded within the electromagnetic action (\ref{em action}) for suitable 
forms of the EFT coupling parameters $f_i(\eta)$. For example,
\begin{itemize}
 \item The model where the EM field couples with a scalar field (generally the inflaton field), in particular the action is given by,
 \begin{eqnarray}
  S_{m1} = \int d^4x\sqrt{-g}\left[f(\phi)F_{\mu\nu}F^{\mu\nu}\right]
  \label{new action}
 \end{eqnarray}
where $\phi$ is the scalar field under consideration and $f(\phi)$ is the conformal breaking coupling function. The above magnetogenesis model 
without (i.e instantaneous reheating) or with reheating phase have been explored earlier in \cite{Demozzi:2009fu,Haque:2020bip,Kobayashi:2019uqs} 
where the coupling function is 
taken as $f(\phi(\eta)) = \left(a(\eta)/a_f\right)^n$, with $\phi = \phi(\eta)$ being determined by the background evolution of the scalar field 
and $a(\eta)$ is the scale factor of the FRW metric. The possible implications of an elongated reheating phase 
have been discussed in such magnetogenesis scenario, 
and moreover the parameter $n$ is constrained for which the model predicts sufficient magnetic strength and also becomes free from 
various problems like the backreaction issue, the strong coupling problem etc \cite{Demozzi:2009fu,Haque:2020bip,Kobayashi:2019uqs}.\\
Most importantly, the action (\ref{new action}) has a direct correspondence with the EFT action given in Eq.(\ref{em action}) when the 
EFT parameters get the following forms:
\begin{eqnarray}
 f_1(\eta) = \left(\frac{a(\eta)}{a_f}\right)^n~~~~~~,~~~~~f_2(\eta) = f_3(\eta) = f_4(\eta) = 0
 \label{correspondence}
\end{eqnarray}
respectively.

\item On contrary to the scalar field coupled magnetogenesis model, let us consider the model where the EM field couples with the background 
spacetime curvature, in particular with the Ricci scalar as well as with the Gauss-Bonnet scalar curvature. The corresponding action is \cite{Bamba:2020qdj},
\begin{eqnarray}
 S_{m2} = \int d^4x\sqrt{-g}\left[-\frac{1}{4}F_{\mu\nu}F^{\mu\nu} + f(R,\mathcal{G})F_{\mu\nu}F^{\mu\nu}\right]
 \label{new action2}
\end{eqnarray}
where the coupling function $f(R,\mathcal{G})$ spoils the conformal invariance of the EM field and has the form as 
$f(R,\mathcal{G}) = \kappa^{2q}\left(R^{q} + \mathcal{G}^{q/2}\right)$, with $q$ being the model parameter. 
This magnetogenesis model without (i.e instantaneous reheating) and with reheating phase have been explored 
in \cite{Bamba:2020qdj} where an elongated reheating phase seems to have a considerable impact on the magnetic field evolution, and in turn it predicts 
sufficient magnetic strength for suitable range of the reheating EoS. Moreover 
the above form of $f(R,\mathcal{G})$ does not lead to scale invariant magnetic or electric power spectrum for the possible values of $q$.\\
Once again, the action $S_{m2}$ can be embedded within the EFT action (\ref{em action}), provided the EFT parameters are considered to have the 
following forms,
\begin{eqnarray}
 f_1(\eta)&=&-\frac{1}{4} + \kappa^{2q}\left(R^{q} + \mathcal{G}^{q/2}\right)\nonumber\\
 f_2(\eta)&=&f_3(\eta) = f_4(\eta) = 0
 \label{correspondence2}
\end{eqnarray}
respectively, where $R(\eta) = \frac{6}{a^2}\left(\mathcal{H}' + \mathcal{H}^2\right)$ and $\mathcal{G}(\eta) = \frac{24}{a^4}\mathcal{H}^2\mathcal{H}'$, 
with $\mathcal{H}$ represents the conformal Hubble parameter in the FRW metric.

\item In both the above magnetogenesis models, the equivalent EFT parameters like $f_3(\eta)$ and $f_4(\eta)$ become zero, mainly due to the absence of 
the parity violating terms in the EM field Lagrangian. However, on contrary, we can consider certain magnetogenesis models where the parity violating terms 
are indeed present in the EM Lagrangian and the equivalent $f_3(\eta)$ or $f_4(\eta)$ (or both) are non-zero. One of such models is explored by the 
authors of \cite{Kushwaha:2020nfa}, where the EM field is considered to couple with the background Riemann tensor, particularly the action is,
\begin{eqnarray}
 S_{m3} = \int d^4x\sqrt{-g}\left[-\frac{1}{4}F_{\mu\nu}F^{\mu\nu} - \frac{\sigma}{M^2}\tilde{R}^{\mu\nu\alpha\beta}F_{\mu\nu}F_{\alpha\beta}\right]
 \label{new action3}
\end{eqnarray}
where $\sigma$ is a model parameter, $R_{\rho\sigma}^{\alpha\beta}$ is the Riemann tensor and its dual is 
$\tilde{R}^{\mu\nu\alpha\beta} = \frac{1}{2}\epsilon^{\mu\nu\rho\sigma}R_{\rho\sigma}^{\alpha\beta}$. 
Due to the presence of the dual tensor in the Lagrangian, the two polarization modes of the EM field evolve differently, 
which in turn leads to the generation of helical magnetic field. The non-minimal coupling of the EM field to the Riemann tensor generates sufficient 
primordial helical magnetic fields at large scales, and the model is also free from the backreaction, strong coupling problems.\\
In the case of background FRW spacetime (with $a(\eta)$ being the scale factor), the action $S_{m3}$ turns out to be,
\begin{eqnarray}
 S_{m3} = \int d^4x\sqrt{-g}\left[-\frac{1}{4}F_{\mu\nu}F^{\mu\nu} + \frac{\sigma}{2M^2}\left(\frac{a'^2}{a^2} + \frac{a''}{a}\right)
 \epsilon^{ijk}F^{0}_{i}F_{jk}\right]~~.
 \label{new action3 modified}
\end{eqnarray}
Thereby interestingly, the $S_{m3}$ can be thought as equivalent with the EFT action given in Eq.(\ref{em action}) 
if the the EFT parameters have the forms like,
\begin{eqnarray}
 f_1(\eta)&=&-\frac{1}{4}~~~~~~~~~~,~~~~~~~~~~~f_2(\eta) = 0~~,\nonumber\\
 f_3(\eta)&=&\frac{\sigma}{2M^2}\left(\frac{a'^2}{a^2} + \frac{a''}{a}\right)~~~~,~~~~f_4(\eta) = 0
 \label{correspondence3}
\end{eqnarray}
respectively. Here it may be mentioned that in \cite{Kushwaha:2020nfa}, the authors considered an $instantaneous~reheating$ mechanism 
in regard to the magnetogenesis model (\ref{new action3}), however one may expect that the inclusion of a reheating phase with 
non-zero e-fold number will enhance the magnetic strength at present epoch compared to the instantaneous reheating case.

\item As a well known one, we consider the Turner-Widrow model \cite{Turner:1987bw} where the action is given by,
\begin{eqnarray}
 S_{m4} = \int d^4x\sqrt{-g}\left[-\frac{1}{4}F_{\mu\nu}F^{\mu\nu} + \frac{1}{12}RA^2\right]
 \label{new action4}
\end{eqnarray}
where $R$ is the background Ricci scalar. The presence of the effective potential term for the vector field (i.e $V\left(A_{\mu}A^{\mu}\right)$) 
in the above action spoils the U(1) invariance of the electromagnetic field, due to which the massive vector field 
gets three physical degrees of freedom: two of them are usual transverse modes and the other one is the longitudinal mode. 
The longitudinal mode with momentum $p^2 > R/6$ appears as a ghost field in the model, i.e a field with negative kinetic
energy \cite{Himmetoglu:2009qi,Himmetoglu:2008zp,Himmetoglu:2008hx,Karciauskas:2010as}. 
However on contrary, in the present work, it is the kinetic term of the electromagnetic field that gets coupled through the EFT parameters, 
and hence there is no potential term of the electromagnetic field appearing in the action. Thus the U(1) invariance is preserved in the
present model, and consequently the longitudinal mode is absent. This clearly indicates that the action $S_{m4}$ is not equivalent with the EFT 
action (\ref{em action}) considered in this work.

\end{itemize}

\end{document}